\documentclass[preprint,authoryear,12pt]{elsarticle-}

\usepackage{amsmath,amssymb,amsfonts,amsthm}
\usepackage{graphicx,tabls}
\usepackage{natbib}
\usepackage[margin=0.5cm]{caption}
\usepackage{color}
\usepackage{hyperref}

\newtheorem{proposition}{Proposition}[section]

\newtheorem{remark}{Remark}[section]
\newtheorem{algorithm}{Algorithm}[section]

\newcommand{\bm}[1]{\mbox{\boldmath{$#1$}}}
\newcommand{\dd}{\hbox{d}}

\begin{document}

\begin{frontmatter}

\title{EM algorithms for estimating the Bernstein copula}

\author{Xiaoling Dou\corref{cor1}\fnref{label1}}
\ead{xiaoling@ism.ac.jp}
\cortext[cor1]{Corresponding author.}
\address[label1]{
The Institute of Statistical Mathematics,
10-3 Midoricho, Tachikawa, Tokyo 190-8562, Japan}

\author{Satoshi Kuriki\fnref{label1}}
\ead{kuriki@ism.ac.jp}

\author{Gwo Dong Lin\fnref{label2}}
\ead{gdlin@stat.sinica.edu.tw}
\address[label2]{
Institute of Statistical Science, Academia Sinica,
Taipei 11529, Taiwan, R.O.C.}

\author{Donald Richards\fnref{label1,label3}}
\ead{richards@stat.psu.edu}
\address[label3]{
Department of Statistics, Penn State University,
University Park, PA 16802, U.S.A.}

\begin{abstract}
A method that uses order statistics to construct multivariate distributions with fixed marginals and which utilizes a representation of the Bernstein copula in terms of a finite mixture distribution is proposed.  Expectation-maximization (EM) algorithms to estimate the Bernstein copula are proposed, and a local convergence property is proved.  Moreover, asymptotic properties of the proposed  semiparametric estimators are provided.  Illustrative examples are presented using three real data sets and a 3-dimensional simulated data set.  These studies show that the Bernstein copula is able to  represent various distributions flexibly and that the proposed EM algorithms work well for such data.
\end{abstract}

\end{frontmatter}

\section{Introduction} 

We consider a far-reaching idea of \cite{Baker08}, who proposed a simple and intuitive method using order statistics for constructing multivariate distributions with given marginals.  Baker's idea in the case of bivariate distributions can be stated as follows: Let $X_1,\ldots,X_m$ and $Y_1,\ldots,Y_n$ be independent random samples from cumulative distribution functions $F$ and $G$, respectively, where $F$ and $G$ can be continuous or discrete.  By sorting the two samples, we obtain the corresponding order statistics $X_{(1)}\le\cdots\le X_{(m)}$ and $Y_{(1)}\le\cdots\le Y_{(n)}$, respectively.  Furthermore, independently of $X_1,\ldots,X_m$ and $Y_1,\ldots,Y_n$, we choose $K$ and $L$, uniformly distributed random numbers from the sets $\{1,\ldots, m\}$ and $\{1,\ldots, n\}$, respectively; then it is straightforward to show that the respective marginal distributions of $X_{(K)}$ and $Y_{(L)}$ are $F$ and $G$, the same as those of $X_k$ and $Y_l$.  The joint distribution of $(X_{(K)}, Y_{(L)})$, the pair of the $K$th and $L$th smallest order statistics from the individual $X$- and $Y$-samples, respectively, is called {\it Baker's bivariate distribution}, and Baker's multivariate distribution can be defined in a similar way.

According to the above construction, we see that Baker's bivariate distribution is parameterized by an $m \times n$ matrix parameter $R=(r_{k,l})$, where
\[
 r_{k,l} = \Pr(K=k,L=l), \ \ 1\le k\le m,\ 1\le l\le n.
\]
Because the marginal distributions of $K$ and $L$ are both uniform, we find that $R$ satisfies the conditions, 
\begin{align}
 \sum_{l=1}^n r_{k,l}=\frac{1}{m}, \quad
 \sum_{k=1}^m r_{k,l}=\frac{1}{n}, \quad
 r_{k,l} \ge 0
\quad \mbox{for}\ 1\le k\le m,\ 1\le l\le n.
\label{sumr}
\end{align}
If $r_{k,l} = 1/(mn) $ for all $k$ and $l$, that is, if $K$ and $L$ are independent then $X_{(K)}$ and $Y_{(L)}$ are also independent.  Otherwise, $K$ and $L$ are not independent and
$(X_{(K)},Y_{(L)})$ is a correlated bivariate random variable.

Let $F_{k:m}$ and $G_{l:n}$ be the marginal distribution functions of the order statistics $X_{(k)}$ and $Y_{(l)}$, respectively.  Then, Baker's distribution is a finite mixture distribution of $mn$-components with distribution function
\begin{equation}
\label{H}
 H(x,y;R) = \Pr\bigl(X_{(K)}\le x,\,Y_{(L)}\le y\bigr) =
 \sum_{k=1}^m \sum_{l=1}^n r_{k,l} F_{k:m}(x) G_{l:n}(y).
\end{equation} 
It is well-known that the distribution functions of order statistics can be described in terms of the Bernstein polynomials; see \cite{Baker08}.  Let the Bernstein polynomial and its cumulative integral be
\[
 b_{k,n}(u) = \binom{n}{k} u^{k} (1-u)^{n-k}, \quad
 B_{k,n}(u) = \int_0^u b_{k,n}(t) dt, \quad u\in [0,1],
\]
respectively.  Then, the distribution functions $F_{k:m}$ and $G_{l:n}$ can be expressed as
$F_{k:m}(x)=m B_{k-1,m-1}(F(x))$ and $G_{l:n}(y)=n B_{l-1,n-1}(G(y))$ \citep[Eq. (1)]{Hwang-Lin84}.  Substituting these results into (\ref{H}), we have
\begin{equation}
\label{H2}
 H(x,y;R) = C(F(x),G(y);R),
\end{equation}
where
\begin{equation}
\label{C}
C(u,v;R) = m n\sum_{k=1}^m\sum_{l=1}^n r_{k,l}
 B_{k-1,m-1}(u)B_{l-1,n-1}(v), \quad (u,v)\in[0,1]^2.
\end{equation}

We now recall that a $2$-dimensional copula is an arbitrary bivariate distribution function
on $[0,1]^2$ whose marginals are the uniform distribution on $[0,1]$.  The importance of copulas in the study of multivariate distributions stems from Sklar's theorem: Any multivariate distribution function can be represented by a copula evaluated at the corresponding marginal distribution functions \citep{Joe01, Nelsen06}. 

In fact, the function $C(u,v; R)$ in (\ref{C}) is a copula, and therefore (\ref{H2})
expresses Baker's distribution explicitly in terms of a copula with arguments $F$ and $G$ and 
parameter $R$; hence, (\ref{H2}) provides an explicit formulation of Sklar's theorem for
Baker's distribution. In this paper, we call $C(u,v; R)$ the {\it Bernstein copula} \citep{Sancetta-Satchell04}.  

It is also well-known that copulas are useful for describing multivariate distributions, and many copulas have been proposed for that purpose.  Since a copula is a distribution function, we refer to its density as a {\it copula density}.

Among the class of copulas, the Bernstein copula has two remarkable features.  First, because of the Weierstrass approximation theorem, any $2$-dimensional copula can be approximated uniformly on $[0,1]^2$ by the Bernstein copula density
\begin{equation}
\label{Bcopula}
c(u,v;R)=\frac{\partial^2}{\partial u\partial v}C(u,v;R),
\end{equation}
when $m$ and $n$ are sufficiently large \citep{Kingsley51}.  Therefore, any continuous bivariate density function can be approximated by the density arising from the Bernstein copula.  Taking advantage of this result, \cite{Sancetta-Satchell04} proposed an empirical Bernstein copula density estimator and studied its consistency in mean-square-error, and \citet{Janssen-etal12} derived the almost sure consistency and asymptotic normality properties of the empirical Bernstein copula density estimator.

A second remarkable feature of the Bernstein copula is that it is a finite mixture distribution, as stated in (\ref{H}) or (\ref{C}).  Following the definition of Baker's distribution, it is easy to generate random numbers from the joint distribution of $(X_{(K)},Y_{(L)})$.  Because copulas are used not only for analyzing existing data, but also for making predictions through Monte Carlo simulation, the simplicity of data generation increases the importance of the Bernstein copula for practical applications.  Moreover, the finite mixture nature of the distribution function (\ref{H2}) allows us to apply the expectation-maximization (EM) algorithm to estimate parameters, and we propose such estimation methods in this paper.

We remark that the EM algorithm is a widely-used method for deriving maximum likelihood estimators (MLEs) numerically.  For the purposes of estimating the Bernstein copula, we prefer the EM algorithm to direct methods such as the Newton-Raphson or quasi-Newton methods,
because the EM algorithm has the advantages of being easy to implement, of not requiring the computation of gradients or Hessians, and of being generally stable and not overly sensitive to starting values even when there exist multiple local maxima of the likelihood function \citep{McLachlan-Krishnan08}.

The paper is organized as follows.  In Section \ref{sec:em}, EM algorithms for estimating parameters are proposed in various settings.  For the first EM algorithm, we prove the local convergence of the M-step, and asymptotic properties of the proposed estimators as semiparametric estimators are provided.  In Section \ref{sec:examples}, we provide illustrative examples based on Baker's distribution, and we illustrate behavior of the proposed algorithms using real-world and simulated data.  Further, some additional properties of the proposed algorithms and two topics for future research are discussed in Section \ref{sec:discussion}, and we provide related mathematical details in the Appendix.

\section{EM algorithms based on the pseudo-likelihood function}
\label{sec:em}

As we have noted, Baker's distribution is a finite mixture distribution, and hence EM algorithm methodology can be applied for maximum likelihood estimation \citep{McLachlan-Peel00}.  Throughout this paper, we assume that the marginal distributions $F$ and $G$
have been estimated in advance, and we shall treat them in the subsequent analysis as known functions; this widely-used two-stage estimation procedure is referred to as the {\it semiparametric method} \citep{Genest-etal95, Charpentier-etal07, Kim-etal07, Choros-etal10}.
 
On the basis of a random sample of size $N$ on $(X,Y)$, let $F_N$ and $G_N$ denote the marginal empirical distributions of $X$ and $Y$.  Throughout the paper, we take $F$ and $G$ to be estimated by $NF_N/(N+1)$ and $NG_N/(N+1)$, respectively.  If $f$ and $g$, the corresponding density functions of $F$ and $G$, exist then we estimate them with kernel estimators (see Section \ref{sec:examples}).  The likelihood function with $F$, $G$, $f$ and $g$ replaced by their corresponding estimators is called the {\it pseudo-likelihood function}.

\subsection{The continuous case}
\label{subsec:cont}

In this subsection, we suppose that $X$ and $Y$ are continuous random variables, and that $F$ and $G$ are absolutely continuous with densities $f$ and $g$, respectively.  The density functions of their $k$th and $l$th smallest order statistics, based on random samples of sizes $m$ and $n$ respectively, can be written as
\begin{equation}
\begin{aligned}
\label{f_km}
f_{k:m}(x)=\,&\frac{\dd}{\dd x} F_{k:m}(x)=m b_{k-1,m-1}(F(x)) f(x), \\
g_{l:n}(y)=\,&\frac{\dd}{\dd y} G_{l:n}(y)=n b_{l-1,n-1}(G(y)) g(y). 
\end{aligned}
\end{equation}
It now follows from (\ref{H}) that the density of Baker's bivariate distribution can be written as
\begin{align}
 h(x,y;R)=\sum_{k=1}^m \sum_{l=1}^n r_{k,l} f_{k:m}(x) g_{l:n}(y).
\label{h_xy}
\end{align}
By applying (\ref{Bcopula}) to (\ref{C}), we obtain the copula density
\[
 c(u,v;R) = m n\sum_{k=1}^m\sum_{l=1}^n r_{k,l}
 b_{k-1,m-1}(u)b_{l-1,n-1}(v),
\]
and then it follows from Sklar's theorem that the density (\ref{h_xy}) has an alternative expression, 
\[
 h(x,y;R)=c(F(x),G(y);R) f(x)g(y).
\]

Suppose that an independent, identically distributed (i.i.d.) sample $(x_i, y_i)$, $i=1,\ldots,N$, is obtained from Baker's distribution (\ref{h_xy}).
According to the standard method for estimating a finite mixture distribution,
we introduce a pair of unobserved variables $(K_i,L_i)$
for observation $i$, with probability
$\Pr(K_i=k, L_i=l)=r_{k,l}$, $k \in \{1,\ldots,m\}$, $l \in \{1,\ldots,n\}$,
$i=1,\ldots,N$. 
We also define an $m \times n$ matrix $\tau_i=(\tau_{i,k,l})$
as a dummy variable with elements
\begin{equation}
\tau_{i,k,l}=
\begin{cases}
 1, & \hbox{if } (K_i, L_i)=(k,l), \\
 0, & \hbox{if } (K_i, L_i) \neq (k,l)
\end{cases}
\label{tau}
\end{equation}
$i=1,\ldots,N$.  Note that $\tau_i$ and $(K_i,L_i)$ are one-to-one.
The likelihood for the full data set $(x_i,y_i,\tau_i)$, $i=1,\ldots,N$, is
given by
\begin{align}
 \prod^N_{i=1} \prod_{k=1}^m \prod_{l=1}^n
 \bigl\{ r_{k,l}f_{k:m}(x_i)g_{l:n}(y_i) \bigr\}^{\tau_{i,k,l}}.
\label{Nrfg}
\end{align}
The E-step in the EM algorithm calculates the conditional expectation
of $\tau_{i,k,l}$
given $(x_i,y_i)$, $i=1,\ldots,N$; that is,
\begin{align}
\widehat{\tau}_{i,k,l}
&= E\bigl[ \tau_{i,k,l} \,|\, (x_i,y_i)_{1\le i\le N};R \bigr] \nonumber \\
&= \frac{r_{k,l} f_{k:m}(x_i)g_{l:n}(y_i)}
    {h(x_i,y_i;R)} \nonumber \\
&= \frac{r_{k,l} b_{k-1,m-1}(F(x_i)) b_{l-1,n-1}(G(y_i))}
    {c(F(x_i),G(y_i);R)}.
\label{hat.tau}
\end{align}
The M-step maximizes the logarithm of the likelihood (\ref{Nrfg})
with respect to $r_{k,l}$ by assuming $\tau_{i,k,l}= \widehat{\tau}_{i,k,l}$.
The logarithm of the expectation of (\ref{Nrfg}) divided by $N$ is
\begin{align}
\frac{1}{N} \sum^N_{i=1} \sum_{k=1}^m \sum_{l=1}^n
\widehat{\tau}_{i,k,l} \log (r_{k,l}f_{k:m}(x_i) g_{l:n}(y_i))
 =\sum_{k=1}^m \sum_{l=1}^n \bar{\tau}_{k,l} \log r_{k,l} + \mathrm{const.},
\label{elog}
\end{align}
where $\bar{\tau}_{k,l}=\sum^N_{i=1} \widehat{\tau}_{i,k,l}/N$.

Maximizing the function (\ref{elog}) is a convex problem 
which has a unique maximizer $R^{*}=(r_{k,l}^{*})$
because (\ref{elog}) is a proper concave function in $r_{k,l}$ and
the region for $R=(r_{k,l})$ defined by (\ref{sumr}) is convex.
Moreover, if $\bar\tau_{k,l}>0$ for all $k,l$ then the maximizer $R^{*}$
is a (relative) interior point of the region (\ref{sumr});
in that case, the maximizer $R^{*}$ is obtained by
the Lagrange multiplier method under the conditions
$\sum_{l=1}^n r_{k,l}=1/m$, $\sum_{k=1}^m r_{k,l}=1/n$
for all $k$ and $l$.

We introduce Lagrange multipliers $\mu_{k}$ and $\lambda_{l}$, and
proceed to maximize
\[
 L = \sum_{k=1}^m \sum_{l=1}^n \bar{\tau}_{k,l} \log r_{k,l}
    -\sum_{k} \mu_{k}\biggl(\sum_{l}r_{k,l}-\frac{1}{m}\biggr)
    -\sum_{l}\lambda_{l}\biggl(\sum_{k}r_{k,l}-\frac{1}{n}\biggr)
\]
with respect to $r_{k,l}$, $\mu_{k}$ and $\lambda_{l}$.
Then, the maximizers $r_{k,l}^{*}$, $\mu_{k}^{*}$ and $\lambda_{l}^{*}$ are
obtained as the solution of
\begin{align*}
\frac{\partial L}{\partial r_{k,l}} = \ & \frac{\bar{\tau}_{k,l}}{r_{k,l}} -\mu_{k}-\lambda_{l} = 0 \end{align*}
subject to the restrictions in (\ref{sumr}).  

To find $\mu_{k}^{*}$ and $\lambda_{l}^{*}$ satisfying
\begin{align}
 r_{k,l}=\frac{\bar{\tau}_{k,l}}{\mu_{k}+ \lambda_{l}} > 0
\label{rkl}
\end{align}
as well as the restriction (\ref{sumr}), we propose the following procedure:

\begin{algorithm}
\label{alg:m}
\hfill

Step M0: 
Set $\mu^{(0)}_{k}=1/2$ and $t=0$.

Step M1: 
For fixed $\bm{\mu}^{(t)}=\bigl(\mu_1^{(t)},\ldots,\mu_m^{(t)}\bigr)'$,
and for $1\le l \le n$, find $\lambda_l^{(t)}$ numerically as a unique 
solution $\lambda_l$ of
\[
 \sum_{k=1}^m\frac{\bar{\tau}_{k,l}}{\mu_k^{(t)}+\lambda_l}=\frac{1}{n}
\quad\mbox{such that}\ \ \lambda_l>-\min_k\bigl(\mu_k^{(t)}\bigr).
\]

Step M2: 
For fixed $\bm{\lambda}^{(t)}=\bigl(\lambda_1^{(t)},\ldots,\lambda_n^{(t)}\bigr)'$,
and for $1 \le k \le m$,
find $\widetilde\mu_k^{(t)}$ numerically as a unique solution $\widetilde\mu_k$ of
\[
 \sum^n_{l=1}\frac{\bar{\tau}_{k,l}}{\widetilde\mu_k+\lambda_l^{(t)}}=\frac{1}{m}
\quad\mbox{such that}\ \ \widetilde\mu_k>-\min_l\bigl(\lambda_l^{(t)}\bigr).
\]

Step M3:
Let
\[
 \mu_{k}^{(t)} = \widetilde\mu_{k}^{(t)}-\frac{1}{m}
  \biggl( \sum_{k=1}^m \widetilde\mu_{k}^{(t)} -\sum_{k=1}^m \mu^{(0)}_{k} \biggr),
 \quad 1 \le k \le m.
\]

Increase the counter $t$ by $1$, and repeat Steps M1--M3 until (\ref{rkl}) converges.
\end{algorithm}

\begin{remark}
\label{rem:uniqueness}
In Step M1, the equation
$\sum_{k=1}^m \bar{\tau}_{k,l}/(\mu_k^{(t)}+\lambda_l)=1/n$ in $\lambda_l$
has at most $m$ solutions.  The solution of (\ref{rkl}) necessarily satisfies
$\mu_k+\lambda_l>0$ for all $(k,l)$, and hence, a solution $\lambda_l$ can be 
chosen to satisfy $\lambda_l>-\min_k\bigl(\mu_k^{(t)}\bigr)$.
This solution is unique because the function 
$\sum_{k=1}^m \bar{\tau}_{k,l}/(\mu_k^{(t)}+\lambda_l)$ is
monotonically decreasing in $\lambda_l$ and takes the values
$\infty$, as $\lambda_l\downarrow -\min_k\bigl(\mu_k^{(t)}\bigr)$,
and $0$, as $\lambda_l\uparrow \infty$.
Such $\lambda_l$ can be found numerically by the bisection method
\citep{DennisJr-Schnabel96}.
Moreover, similar remarks apply to Step M2.
\end{remark}

\begin{remark}
\label{rem:redundancy}
We note that if $\mu_k$ and $\lambda_l$ are solutions of (\ref{rkl}) then $\mu_k+c$ and $\lambda_l -c$ are also solutions of (\ref{rkl}) for any constant $c$.  Therefore, Step M3 is needed to remove this redundancy.
\end{remark}

The following proposition states that Algorithm \ref{alg:m} converges locally, and a proof is given in the Appendix.  An empirical study suggests that this algorithm also has the global convergence property (see Section \ref{subsec:consomic}), however it remains an open problem to derive a proof of the global convergence property.  

\begin{proposition}
\label{prop:local_convergence}
Suppose that $\bar\tau_{k,l}>0$.
Then Algorithm \ref{alg:m} has the property of locally linear convergence.
That is, there exist positive constants $c$ and $d$ such that if 
$\Vert\bm{\mu}^{(0)}-\bm{\mu}^{*}\Vert\le c$ and 
$\Vert\bm{\lambda}^{(0)}-\bm{\lambda}^{*}\Vert\le d$, then 
the sequences $\bm{\mu}^{(t)}$ and $\bm{\lambda}^{(t)}$ 
$(t=0,1,\ldots)$ generated from Algorithm \ref{alg:m} converge to the solutions
$\bm{\mu}^{*}$ and $\bm{\lambda}^{*}$, respectively.  Moreover, convergence is 
attained with the convergence rates 
$\Vert\bm{\mu}^{(t)}-\bm{\mu}^{*}\Vert = O(\nu^t)$ and 
$\Vert\bm{\lambda}^{(t)}-\bm{\lambda}^{*}\Vert = O(\nu^t)$, as $t\to\infty$, 
for a positive constant $\nu \in (0,1)$, and the constants $c,d$ and $\nu$ 
depend on $(\bm{\mu}^{*},\bm{\lambda}^{*})$.
\end{proposition}

To apply the EM algorithm, we will use as an initial value of $r_{k,l}$ the estimator given by \cite{Sancetta-Satchell04} and \cite{Janssen-etal12} (see Remark \ref{rem:inconsistent}).  Then the EM algorithm is summarized as follows.

\begin{algorithm}
\label{alg:em}
\hfill

Step 0:
Set $r_{k,l}$ equal to $\widetilde r_{k,l}$ in (\ref{inconsistent}).

Step 1:
Find $\widehat{\tau}_{i,k,l}$ by (\ref{hat.tau}) (E-step).

Step 2:
Update $r_{k,l}$ by Algorithm \ref{alg:m}, Steps M0--M3 (M-step).

Repeat Steps 1 and 2 until $\widehat{\tau}_{i,k,l}$ converges.
\end{algorithm}

Note that this algorithm can be extended to
Baker's distributions with three or more variables.

\begin{remark}
In (\ref{hat.tau}), a common factor
$f(x_i) g(y_i)$ is canceled in the numerator and the denominator.
This is reasonable because the EM algorithm should be equivalent to
the one based on the sample $(F(x_i),G(y_i))$, $i=1,\ldots,N$,
having uniform marginals.
\end{remark}

\cite{Genest-etal95} developed an asymptotic theory for semiparametric
estimation of copulas based on the pseudo-likelihood function, and 
\cite{Tsukahara05} later extended that theory to M-estimation.
Their results can be applied to our problem.  Let
\begin{align*}
c_u(u,v;R) &=
 mn \sum_{k=1}^m \sum_{l=1}^n r_{k,l} \frac{\dd}{\dd u} b_{k-1,m-1}(u) b_{l-1,n-1}(v), \\
c_v(u,v;R) &=
 mn \sum_{k=1}^m \sum_{l=1}^n r_{k,l} b_{k-1,m-1}(u) \frac{\dd}{\dd v} b_{l-1,n-1}(v)
\end{align*}
be the derivatives of $c(u,v)$ with respect to $u$ and $v$.
To calculate these derivatives, we can use the formula 
\[
 \frac{\dd}{\dd u}b_{k,n}(u)=n\{b_{k-1,n-1}(u)-b_{k,n-1}(u)\},
\]
where we set $b_{-1, n-1}(u) = b_{n, n-1}(u) \equiv 0$ to initialize the recurrence relation.

\begin{proposition}
\label{prop:sigma_hat}
Suppose that the true value of the parameter $R=(r_{k,l})$ in (\ref{sumr})
satisfies $r_{k,l}>0$.  As $N \to \infty$ the MLE,
$\widehat R$, of $R$ is an asymptotically normally $\sqrt{N}$-consistent
estimator.  

Further, a consistent estimator of
$N\mathrm{Var}(\widehat R) = \Sigma = (\sigma_{(k,l),(k',l')})$, an 
$mn\times mn$ matrix with lexicographic index $(k,l)$, 
is given by $\widehat\Sigma=B^+ S B^+$,
where $B$ and $S$ are the sample covariance matrices
of the $mn\times 1$ pseudo-observation vectors
$\bm{u}_i=(u_{i,(k,l)})$ and $\bm{v}_i=(v_{i,(k,l)})$ defined by
\begin{align*}
u_{i,(k,l)} = \
& \frac{mn b_{k-1,m-1}(F(x_i)) b_{l-1,n-1}(G(y_i))}{c(F(x_i),G(y_i);\widehat R)}, \\
v_{i,(k,l)} = \
& u_{i,(k,l)} \\
& -\frac{mn}{N}\sum_{j:x_i\le x_j}
   \frac{b_{k-1,m-1}(F(x_j)) b_{l-1,n-1}(G(y_j)) c_u(F(x_j),G(y_j);\widehat R)}
        {c(F(x_j),G(y_j);\widehat R)^2} \\
& -\frac{mn}{N}\sum_{j:y_i\le y_j}
   \frac{b_{k-1,m-1}(F(x_j)) b_{l-1,n-1}(G(y_j)) c_v(F(x_j),G(y_j);\widehat R)}
        {c(F(x_j),G(y_j);\widehat R)^2},
\end{align*}
$i=1,\ldots,N$, respectively, and 
$B^+$ is the Moore-Penrose pseudo-inverse matrix of $B$.
\end{proposition}

Note that the matrix $B$ is the observed Fisher information matrix
when the marginals $F$ and $G$ are known.
As with many semiparametric estimators, 
$\widehat R$ is not efficient in the sense that
its asymptotic variance is larger than that given by the Fisher
information matrix when the marginals are known 
(unless $r_{k,l}\equiv 1/(mn) $, i.e., $c(u,v;R)=1$); see
\cite{Genest-Werker02} regarding the inefficiency of
the Farlie-Gumbel-Morgenstern (FGM) copula estimator.

Once the estimator $\widehat R = (\widehat r_{k,l})$ has been obtained 
by Algorithms \ref{alg:m} and \ref{alg:em},
the estimate of $h(x,y;R)$ for a fixed point $(x,y)$ is given by 
\[
h\bigl(x,y;\widehat R\bigr)=\sum_{k=1}^m \sum_{l=1}^n
 \widehat r_{k,l} f_{k:m}(x) g_{l:n}(y)
\]
and its asymptotic variance is evaluated as
\[
 \mathrm{Var}\big(h (x,y;\widehat R )\big) \approx
 \frac{1}{N} \sum_{k,k'=1}^m \sum_{l,l'=1}^n
 f_{k:m}(x) f_{k':m}(x) g_{l:n}(y) g_{l':n}(y) \widehat\sigma_{(k,l),(k',l')},
\]
where $\widehat\sigma_{(k,l),(k',l')}$ is the $((k,l),(k',l'))$th element of
an $mn\times mn$ matrix $\widehat\Sigma=(\widehat\sigma_{(k,l),(k',l')})$.

\begin{remark}
\label{rem:inconsistent}

\cite{Sancetta-Satchell04} and \cite{Janssen-etal12}
proposed estimating $r_{k,l}$ by
\begin{equation}
 \widetilde r_{k,l}= \#\biggl\{ i \ \Big|\ %
 \frac{k-1}{m}<\frac{N}{N+1}F_N(x_i)\le\frac{k}{m},\ %
 \frac{l-1}{n}<\frac{N}{N+1}G_N(y_i)\le\frac{l}{n} \biggr\} \Big/ N
\label{inconsistent}
\end{equation}
as $m, n \to \infty$, where the sample size $N \to \infty$ also.  
For the case in which $m$ and $n$ are fixed, this estimator is inconsistent in our setting 
because
\begin{align*}
\lim_{N\to\infty} E\bigl[\widetilde r_{k,l}\bigr] =
\ & C\Bigl(\frac{k}{m},\frac{l}{n};R\Bigr)
- C\Bigl(\frac{k}{m},\frac{l-1}{n};R\Bigr) \\
&-C\Bigl(\frac{k-1}{m},\frac{l}{n};R\Bigr)
+ C\Bigl(\frac{k-1}{m},\frac{l-1}{n};R\Bigr)
\end{align*}
is not equal to $r_{k,l}$, in general.
\end{remark}

\subsection{The discrete case}

The EM algorithm in Section \ref{subsec:cont} is also applicable
for the case in which both $F$ and $G$ are discrete distributions.
Suppose that $F$ and $G$ are supported on discrete sets $A$ and $B$, 
respectively.  For simplicity, we suppose that $A$ and $B$ are finite 
sets, with cardinalities $|A|$ and $|B|$, respectively.  Then the data 
$(x_i,y_i)$, $i=1,\ldots,N$, can be represented by an 
$|A|\times |B|$ ordered categorical table $(N_{a,b})$, where
\[
 N_{a,b} = \#\bigl\{i\in\{1,\ldots,N\} \mid (x_i,y_i)=(a, b)\bigr\},
 \quad a \in A,\ b \in B.
\]
In this subsection, we modify the EM algorithm of Section \ref{subsec:cont}
so that Baker's distribution (\ref{H}) can be applied to
the data $(N_{a,b})_{a\in A,\,b\in B}$.

The probability functions of $X$ and $Y$ are
$f(a)=\Pr(X=a)=F(a)-F(a-)$, $a \in A$ and
$g(b)=\Pr(Y=b)=G(b)-G(b-)$, $b \in B$, respectively.
The probability functions of $X_{(k)}$ and $Y_{(l)}$, 
the $k$th and $l$th order statistics, are $f_{k:m}$ 
and $g_{l:n}$, respectively, where
\begin{equation}
\begin{aligned}
\label{g_b}
f_{k:m}(a) \
   & = F_{k:m}(a)-F_{k:m}(a-) \\
   & = m\{ B_{k-1,m-1}(F(a))-B_{k-1,m-1}(F(a-)) \}, \\
g_{l:n}(b)
   & = G_{l:n}(b)-G_{l:n}(b-) \\
   & = n\{ B_{l-1,n-1}(G(b))-B_{l-1,n-1}(G(b-)) \}. 
\end{aligned}
\end{equation}
Using these results, we obtain the joint probability function of 
$(X,Y)$ in the form 
\[
 h(a,b;R) = \Pr(X=a,Y=b)
 = \sum_{k=1}^m \sum_{l=1}^n r_{k,l} f_{k:m}(a) g_{l:n}(b).
\]

We introduce a dummy variable
$\eta_{a,b,k,l}=\sum_{i:(x_i,y_i)=(a,b)}\tau_{i,k,l}$
with $\tau_{i,k,l}$ defined in (\ref{tau}).
The likelihood for the full data (\ref{Nrfg}) is rewritten as
\[
 \prod_{a\in A} \prod_{b\in B} \prod_{k=1}^m \prod_{l=1}^n
 \bigl\{ r_{k,l}f_{k:m}(a)g_{l:n}(b) \bigr\}^{\eta_{a,b,k,l}}.
\]
The E-step for updating $\eta_{a,b,k,l}$ becomes
\[
\widehat{\eta}_{a,b,k,l}
 = E\bigl[ \eta_{a,b,k,l} \,|\, (N_{a,b})_{a\in A,\,b\in B}; R \bigr]
 = \frac{N_{a,b} r_{k,l} f_{k:m}(a)g_{l:n}(b)}
        {\sum_{k=1}^m\sum_{l=1}^n r_{k,l}f_{k:m}(a)g_{l:n}(b)}.
\]
By letting $\bar\tau_{k,l}=\sum_{a\in A}\sum_{b\in B}\widehat\eta_{a,b,k,l}/N $,
the M-step is obtained in the same form in Section \ref{subsec:cont},
which is Step 2 of Algorithm \ref{alg:em}.

\subsection{The mixed case}

We can also resolve by the same approach the case in which one variable 
is continuous and the other is discrete. Suppose that
$X$ is continuous with density function $f$ and $Y$ is discrete with 
distribution function $G$.  Then, the density function of $X_{(k)}$
and the probability function of $Y_{(l)}$ are given by
$f_{k:m}$ in (\ref{f_km}) and $g_{l:n}$ in (\ref{g_b}), respectively.
The joint density function becomes
\begin{align*}
 h(x,&b;R) \\
\ & = \frac{\Pr(X\in (x,x+\dd x),Y=b)}{\dd x} \\
 & = \sum_{k=1}^m \sum_{l=1}^n r_{k,l} f_{k:m}(x) g_{l:n}(b) \\
 & = mn \sum^m_{k=1}\sum^n_{l=1} r_{k,l} b_{k-1,m-1}(F(x)) f(x)
       \{ B_{l-1,n-1}(G(b)) - B_{l-1,n-1}(G(b-))\}.
\end{align*}
The E-step is the updating rule, 
\begin{align*}
\widehat{\tau}_{i,k,l} \
& = E\bigl[\tau_{i,k,l} \,|\, (x_i,y_i)_{1\le i\le N};R \bigr] \\
& = \frac{r_{k,l} f_{k:m}(x_i)g_{l:n}(y_i)}{h(x_i,y_i;R)} \\
& = \frac{r_{k,l} b_{k-1,m-1}(F(x_i))\{ B_{l-1,n-1}(G(y_i))-B_{l-1,n-1}(G(y_i-))\}}
        {\sum^m_{k=1} \sum^n_{l=1} r_{k,l} b_{k-1,m-1}(F(x_i))
         \{ B_{l-1,n-1} (G(y_i)) - B_{l-1,n-1}(G(y_i-))\}},
\end{align*}
and the M-step remains unchanged.

\subsection{The case in which $R$ is parameterized}
\label{subsec:parametric}

If $R=(r_{k,l})$ satisfying (\ref{sumr}) is parameterized by
a lower-dimensional parameter $\theta$ as $r_{k,l}=r_{k,l}(\theta)$ 
then the estimation becomes simpler.
For the case in which $m=n$, for instance, \cite{Baker08} discussed
a subclass of bivariate distributions with a distribution function
\begin{align}
\label{Hpm}
H^{\pm}(x,y;q,n)
 &= (1-q)F(x)G(y) +q H_n^{\pm}(x,y) \nonumber \\
 &= (1-q)F(x)G(y) +q C_n^{\pm}(F(x),G(y)), \quad 0 \le q \le 1, 
\end{align}
where
\begin{align*}
H_n^{+}(x,y) &=\frac{1}{n}\sum^n_{k=1} F_{k:n}(x) G_{k:n}(y)
 = C_n^{+}(F(x),G(y)), \\
C_n^{+}(u,v) &= n\sum^n_{k=1} B_{k-1,n-1}(u) B_{k-1,n-1}(v),
\end{align*}
and
\begin{align*}
H_n^{-}(x,y) &=\frac{1}{n}\sum^n_{k=1} F_{k:n}(x) G_{n-k+1:n}(y)
 = C_n^{-}(F(x),G(y)), \\
C_n^{-}(u,v) &= n\sum^n_{k=1} B_{k-1,n-1}(u) B_{n-k,n-1}(v).
\end{align*}
The densities of $H_n^\pm$ and $C_n^\pm$, if they exist, are denoted by
$h_n^\pm$ and $c_n^\pm$.  The functions $H_n^{+}(x,y)$ and $H_n^{-}(x,y)$ 
correspond, respectively, to the largest positive and smallest negative 
correlation cases among Baker's distributions with $m=n$.  Moreover, the
rank correlation of $H_n^\pm$ is $\pm (n-1)/(n+1)$.

The function $H^{\pm}(x,y;q,n)$ is Baker's distribution (\ref{H2}) with
\[
 r_{k,l}=
\begin{cases}
  (1-q)/n^2 + q\delta_{k,l}/n & (\mbox{for }H^{+}), \\
  (1-q)/n^2 + q\delta_{k,n-l+1}/n & (\mbox{for }H^{-}),
\end{cases}
 \qquad 1\le k,l\le n,
\]
where $\delta_{k,l}$ denotes Kronecker's delta.  The term 
$r_{k,l}$ is parameterized by the scalar parameter $q$ which 
adjusts the degree of independence between $X$ and $Y$.  Indeed, 
if $q=0$ then $X$ and $Y$ are independent; and if $q>0$ then 
$X$ and $Y$ are positively (respectively, negatively) correlated
for the distribution $H^+$ (respectively, $H^-$).
These models are expected to represent highly correlated distributions
with fewer parameters than the original Baker's distribution.

Baker's distribution was originally proposed as an extension to the FGM
distribution with the limitation that its correlation does not exceed $1/3$
for the case of continuous marginals \citep{Schucany-etal78}.  
Hence, the range of correlation of Baker's distribution has gathered attention 
and the extreme correlation cases, $H^{\pm}_n(x,y)$, are well-studied.

For the distribution $H_n^{+}(x,y)$, 
\cite{Lin-Huang10} investigated convergence conditions and the convergence
rate, as $n \to \infty$, of the correlation converging to 
the maximum correlation of the Fr\'echet--Hoeffding upper bound.  
\cite{Dou-etal13} 
proved the TP${}_2$ property and derived the limiting distribution of
$(X,U)$, $U=\sqrt{n}(F(X)-G(Y))$,
where $(X,Y)$ are distributed as $H^{+}_n(x,y)$.
\cite{Huang-etal13} proved that the copula $C_n^+(u,v)$
in the largest correlation case with $u,v$ fixed
is a non-decreasing function of $n$.

In modeling the joint distribution functions, 
\cite{Baker08} chose the parameter $q$
by minimizing the negative log-likelihood and the Kolmogorov-Smirnov statistic
for a given set of values of $n$.
Here, we treat $n$ as an integer-valued parameter to be estimated 
and, as an alternative, 
we propose an EM algorithm below to estimate the parameters $(q,n)$
simultaneously.
Suppose that an i.i.d.\ sample $(x_i,y_i)$, $i=1,\ldots,N$, is obtained
from the continuous distribution $H_n^+(x,y;q,n)$ with the density
\begin{align*}
 h_n^+(x,y;q,n) \
 & = (1-q) f(x) g(y) + q h_n^+(x,y) \\
 & = \{ 1-q + q c_n^+(F(x),G(y)) \} f(x) g(y).
\end{align*}

\begin{algorithm}
\label{alg:qn}
\hfill

Step 0. Set $(q,n)=(1/2,1)$.

Step 1. E-step:
\begin{align}
\widehat{\tau}_i &
:= \frac{(1-q)f(x_i)g(y_i)}{(1-q)f(x_i)g(y_i) +q h_n^{\pm}(x_i,y_i)}
 = \frac{1-q}{1-q +q c_n^{\pm}(F(x_i),G(y_i))},
\label{E-continuous}
\end{align}
$i=1,\ldots,N$. 

Step 2. M-step:
\[
 q := 1-\frac{1}{N}\sum^N_{i=1} \widehat{\tau}_i, 
\qquad
 n := \mathop{\mathrm{argmax}}_{n\in\mathbb{N}}\,
\sum^N_{i=1}
(1-\widehat{\tau}_i) \log\left( c_n^{\pm}(F(x_i),G(y_i)) \right).
\]

Repeat Steps 1 and 2 until $(q,n)$ converges.
\end{algorithm}

The asymptotic variance of $\widehat{q}$ is evaluated approximately 
as $s/(N \beta^2)$, where $\beta$ and $s$ are the sample
variances of the pseudo-observations $u_i$ and $v_i$ defined by
\begin{align*}
u_i \ & = \frac{-1 + c_n^{\pm}(F(x_i),G(y_i))}
            {1-\widehat q+\widehat q c_n^{\pm}(F(x_i),G(y_i))}\bigg|_{n=\widehat n}, \\
v_i
& = u_i -\frac{\widehat q}{N}\sum_{j:x_i\le x_j}
   \frac{ \{-1 + c_n^{\pm}(F(x_j),G(y_j))\} \frac{\partial}{\partial u}c_n^{\pm}(F(x_j),G(y_j))}
        {\{1-\widehat q+\widehat q c_n^{\pm}(F(x_j),G(y_j))\}^2} \bigg|_{n=\widehat n}\\
& \quad -\frac{\widehat q}{N}\sum_{j:y_i\le y_j}
   \frac{ \{-1 + c_n^{\pm}(F(x_j),G(y_j))\} \frac{\partial}{\partial v}c_n^{\pm}(F(x_j),G(y_j))}
        {\{1-\widehat q+\widehat q c_n^{\pm}(F(x_j),G(y_j))\}^2}
\bigg|_{n=\widehat n},
\end{align*}
$i=1,\ldots,N$.

For the case in which both $X$ and $Y$ are discrete distributions,
the E-step (\ref{E-continuous}) in Algorithm \ref{alg:qn} is replaced by 
\begin{equation}
\label{E-discrete}
\widehat{\tau}_i :=
 \frac{(1-q) f(x_i) g(y_i)}
      {(1-q) f(x_i) g(y_i) + (q/n) \sum_{k=1}^n f_{k:n}(x_i) g_{k:n}(y_i)},
\end{equation}
where
$f(x_i)=F(x_i)-F(x_i-)$, $g(y_i)=G(y_i)-G(y_i-)$, and
$f_{k:n}(x_i)$ and $g_{k:n}(y_i)$ are given in (\ref{g_b}).
If the joint distribution consists of both continuous and discrete variables,
then their respective density (\ref{f_km}) and
probability function (\ref{g_b}) should be used.

\section{Four illustrative examples}
\label{sec:examples}

In this section, we demonstrate how our algorithms perform in practical data analysis
for a wide range of examples.
The results show that the algorithms work well in all the illustrative examples.

\subsection{Consomic mouse data}
\label{subsec:consomic}

The first data set, consisting of measurements of blood 
concentrations of biochemical substances in mice,
is available from \cite{Takada-etal12}.  We apply 
Algorithm \ref{alg:em} for fitting Baker's distribution (\ref{h_xy}) 
with continuous variables.

\begin{figure}[!htb]
\begin{center}
 \includegraphics[width=0.7\linewidth]{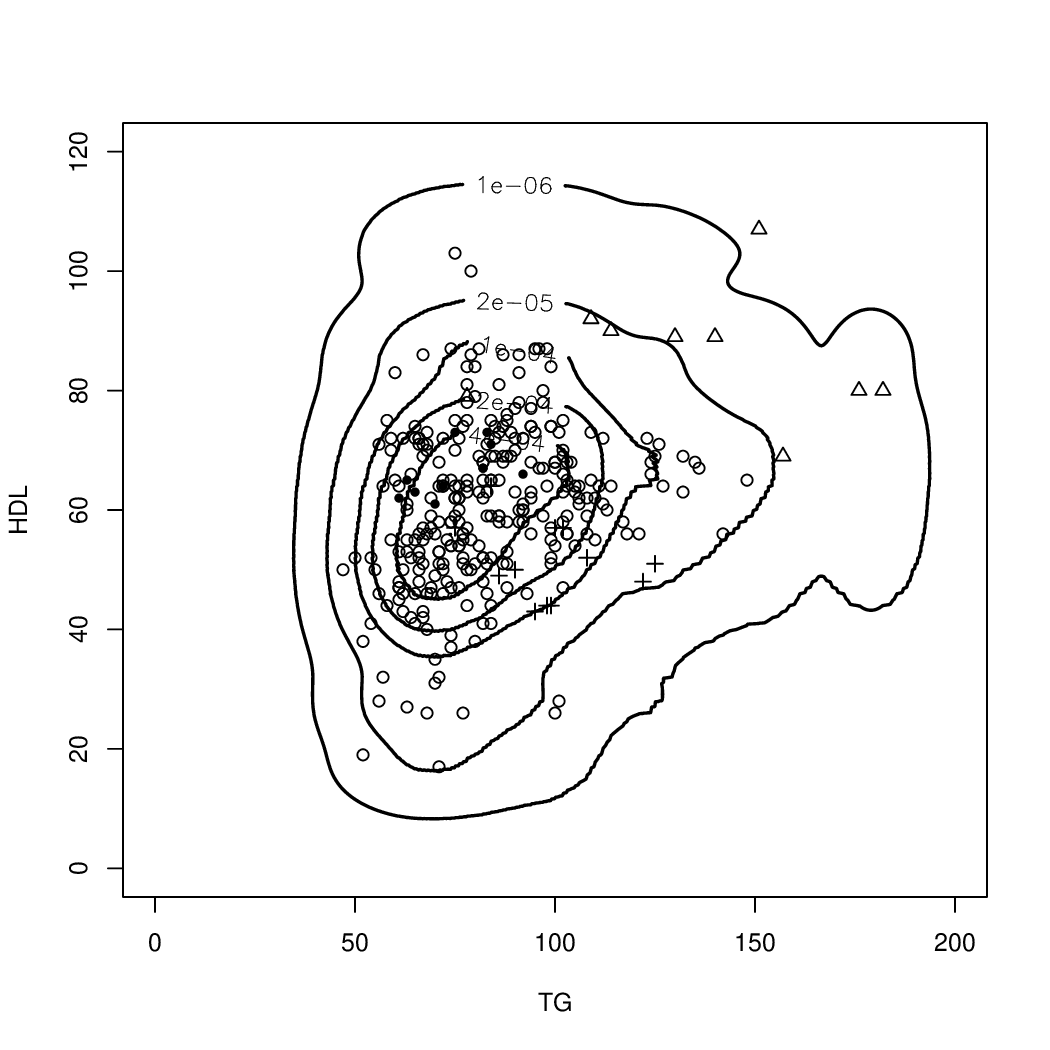}
\end{center}
\vspace{-0.3cm}
\caption{TG and HDL data (female consomic mice) and estimated contour.  \newline
(Dots: B6, Pluses: B6-Chr4MSM, Triangles: MSM, Circles: others.)}
\label{fig:Fmice}
\end{figure}

The data set consists of measurements of
triglycerides (TG) and plasma high-density lipoprotein cholesterol (HDL)
as plotted in Figure \ref{fig:Fmice}.  The variables 
TG and HDL are important indicators of metabolic syndrome and are correlated
with the pathogenesis of cardiovascular disease in humans.
To detect the genes responsible for adiposity, TG and HDL data are taken 
from consomic mouse strains of 314 10-week old females.
A consomic strain is an artificial inbred strain
with one specified chromosome
replaced by another chromosome from a different inbred strain
\citep{Takada-Shiroishi12}.
For example, the label B6-Chr4MSM appearing in Figure \ref{fig:Fmice} means that 
a consomic strain has all chromosomes from the mouse strain C57BL/6 (B6)
except for chromosome 4, which is from the mouse strain MSM/Ms (MSM).
The data are taken from 30 kinds of consomic strains
including pure strains B6 and MSM, and hence are fairly heterogeneous. 

\begin{figure}[!htb]
\begin{center}
\begin{tabular}{ll}
 \includegraphics[width=0.55\linewidth]{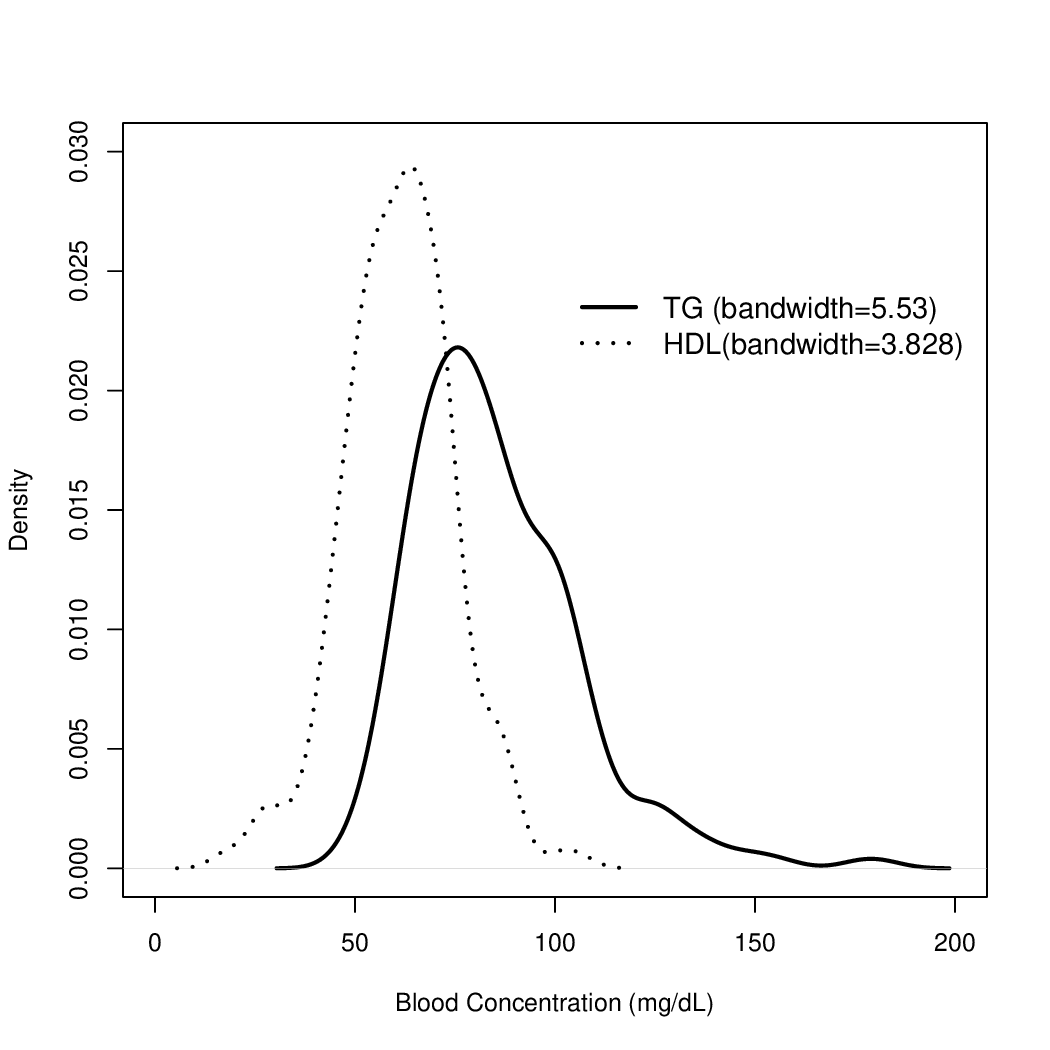} &
 \includegraphics[width=0.55\linewidth]{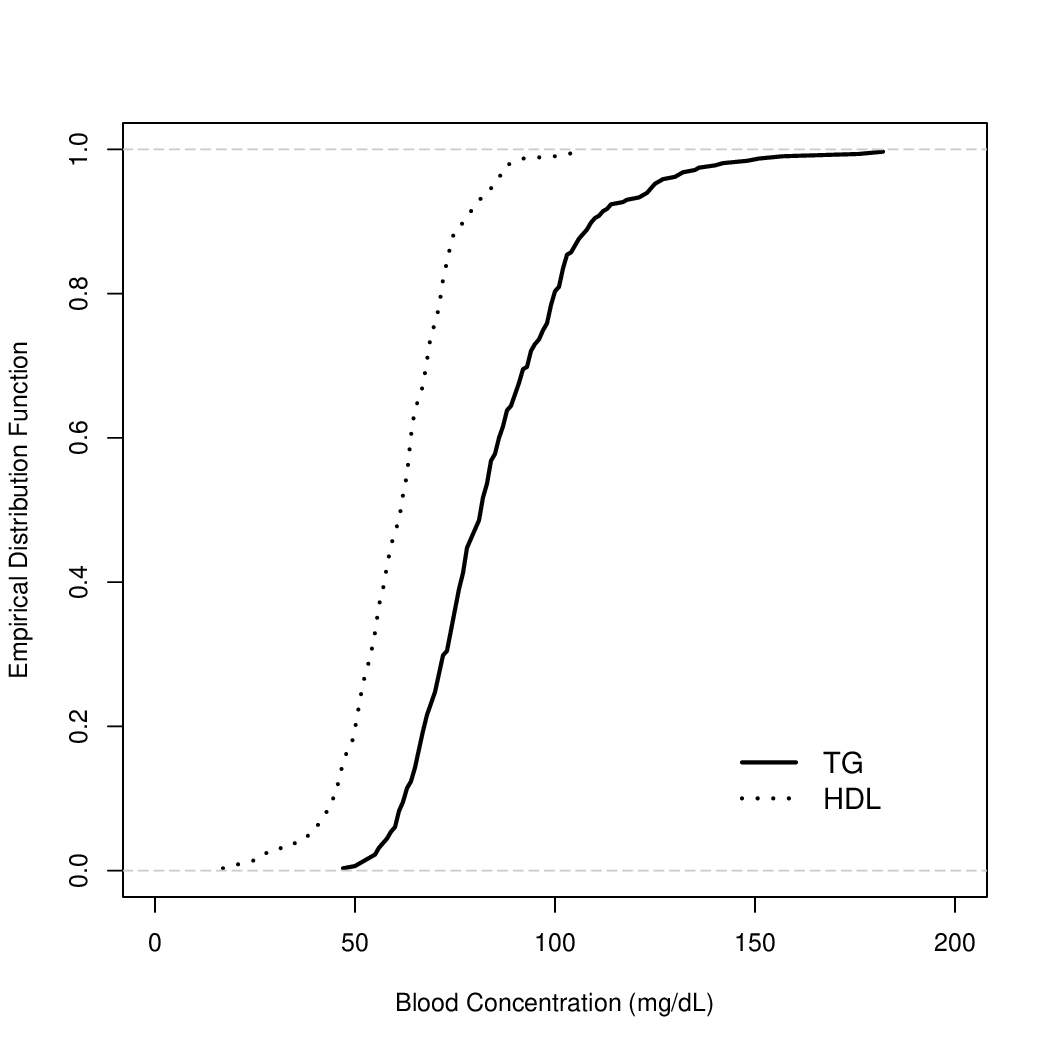} \\
\end{tabular}
\end{center}
\vspace{-0.3cm}
\caption{Estimated marginals of TG and HDL. \newline
 (Left: density functions. Right: cumulative distribution functions.)}
\label{fig:marginals}
\end{figure}

Using the Gaussian kernel estimator,
we first estimate the marginal density functions.  The bandwidths are
selected according to Silverman's ``rule of thumb'' \citep{Silverman86}.
As described in Section \ref{subsec:cont},
we use the empirical distribution functions
to approximate the (cumulative) distribution functions.
The estimated marginal densities and distribution functions are
shown in the left and right panels, respectively, of Figure \ref{fig:marginals}.
Subsequently, we estimate the Bernstein copula density (\ref{h_xy})
with the EM algorithm (Algorithm \ref{alg:em}) for fixed $m$ and $n$.
In the estimation, we determine the matrix size of $R$ by the Akaike 
information criterion (AIC).
From Table \ref{tab:aic}, we find that the AIC attains its minimum
value, $5210.52$,  
when $(m,n)=(2,3)$.
Table \ref{tab:aic} also shows that the cases in which $(m,n)=(2,2)$
and $(m,n)=(2,3)$ have very close AIC values; 
indeed, the estimated contours based on these two cases are very similar.

For the case in which $(m,n)=(2,3)$,
the initial value $\widetilde R$ in (\ref{inconsistent})
and the MLE $\widehat R$ obtained as the limit of sequence
starting from $\widetilde R$ are
\[
\widetilde R=
\begin{pmatrix}
0.232 & 0.137 & 0.118 \\
0.099 & 0.188 & 0.226
\end{pmatrix}
\quad\mbox{and}\quad
\widehat R=
\begin{pmatrix}
0.333 & 0.106 & 0.061 \\
0.000 & 0.227 & 0.273
\end{pmatrix},
\]
respectively.
\begin{table}[htbp]
\caption{AIC for female consomic mouse data. \newline
(The minimum AIC is indicated with a box.)}
\label{tab:aic}
\begin{small}
\begin{tabular}{cccccccccc}
\hline
$m\ \backslash\ n$ & 1 & 2 & 3 & 4 & 5 & 6 & 8 & 10 \\
\hline
1 & 5242.00 & 5242.00 & 5242.00 & 5242.00 & 5242.00 & 5242.00 & 5242.00 & 5242.00 \\
2 & 5242.00 & 5210.57 & \fbox{5210.52} & 5212.15 & 5211.15 & 5210.80 & 5212.67 & 5216.23 \\
3 & 5242.00 & 5212.55 & 5211.94 & 5214.22 & 5215.47 & 5217.64 & 5223.91 & 5230.53 \\
4 & 5242.00 & 5214.56 & 5215.69 & 5219.16 & 5220.33 & 5224.48 & 5234.19 & 5244.29 \\
5 & 5242.00 & 5215.37 & 5218.51 & 5223.65 & 5226.87 & 5232.10 & 5246.20 & 5259.89 \\
6 & 5242.00 & 5216.59 & 5220.58 & 5225.99 & 5231.44 & 5238.67 & 5256.04 & 5273.77 \\
8 & 5242.00 & 5218.77 & 5225.45 & 5233.77 & 5242.13 & 5253.45 & 5277.90 & 5302.69 \\
10& 5242.00 & 5221.55 & 5229.78 & 5241.92 & 5253.58 & 5268.72 & 5300.85 & 5332.22 \\
\hline
\end{tabular}
\end{small}
\end{table}
The consistent estimates of the covariance of
$
(\widehat r_{11},\widehat r_{12},\widehat r_{13},\widehat r_{21},\widehat r_{22},\widehat r_{23})'$ calculated by Proposition \ref{prop:sigma_hat} is
\[
\begin{small}
\begin{pmatrix}
  0.003 & -0.001 & -0.003 & -0.001 &  0.001 & -0.002 \\
 -0.001 &  0.003 &  0.003 & -0.002 &  0.000 &  0.002 \\
 -0.003 &  0.003 &  0.010 & -0.003 & -0.004 &  0.003 \\
 -0.001 & -0.002 & -0.003 &  0.006 &  0.002 & -0.003 \\
  0.001 &  0.000 & -0.004 &  0.002 &  0.003 & -0.002 \\
 -0.002 &  0.002 &  0.003 & -0.003 & -0.002 &  0.005
\end{pmatrix}.
\end{small}
\]
\noindent
A contour plot
of the estimated joint density $h(x,y;\widehat R)$ is shown in Figure \ref{fig:Fmice}.

In fitting models to the data, we checked the convergence of the algorithms
when the starting points vary.
Throughout the estimating procedure, two types of convergence sequences
are generated; one is from Algorithm \ref{alg:m} and the 
other is from Algorithm \ref{alg:em}.
We investigate these convergences for the case in which $(m,n)=(2,3)$ as follows.

For the first type of sequence, we varied the starting points
$\bm{\mu}^{(0)}=\big( \mu_1^{(0)},\mu_2^{(0)} \big) $ as
$(1/2,1/2)$ (as indicated in Algorithm \ref{alg:m}) and
20 pairs of random variables distributed uniformly on $(0,1)^2$.
We find that all of these sequences converge to 
the same limit. 

\begin{figure}[!htb]
\begin{center}
 \includegraphics[width=0.7\linewidth]{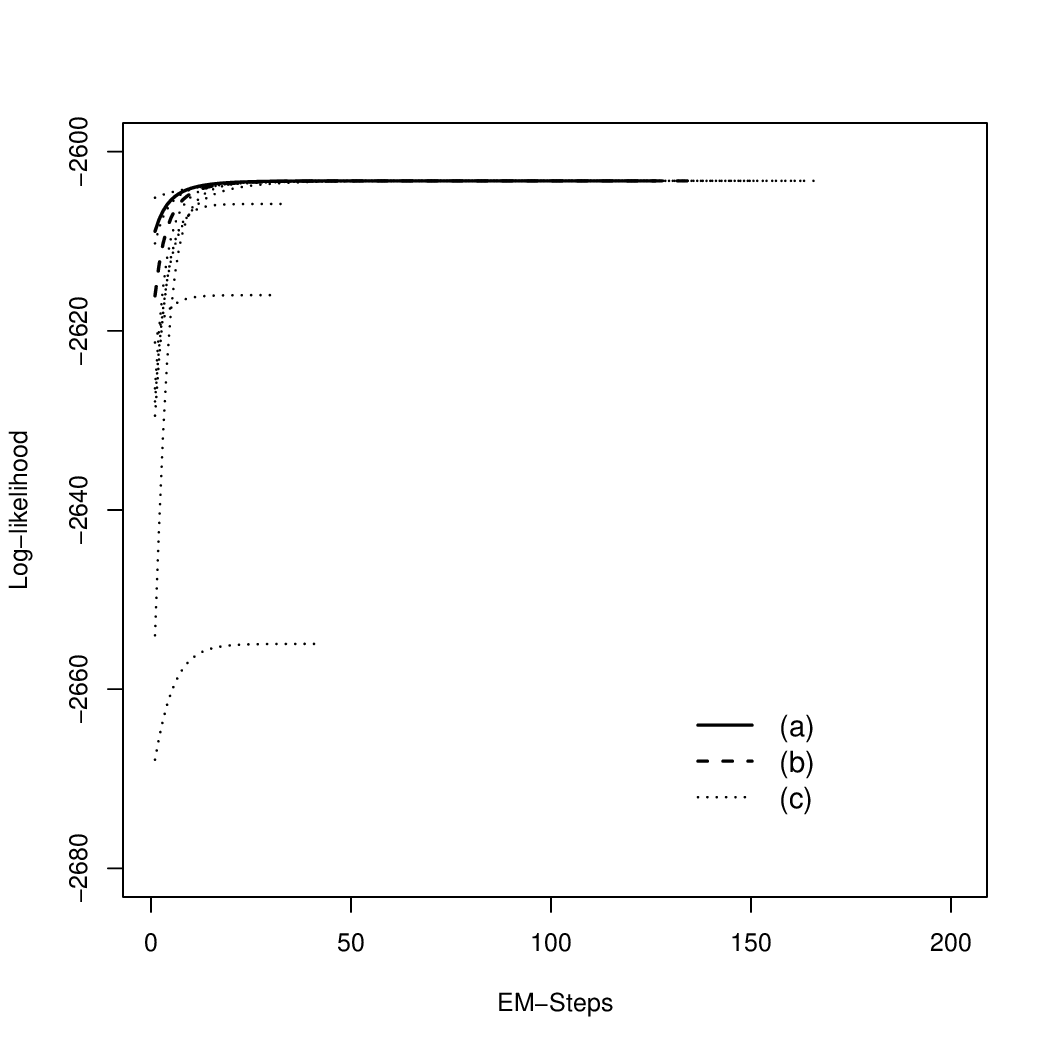}
\end{center}
\vspace{-0.3cm}
\caption{Convergence of the EM algorithm with various starting points. \newline
(a) $R^{(0)}=\widehat R$; 
(b) $r^{(0)}_{k,l} \equiv 1/6$; 
(c) $\big( r^{(0)}_{1,1},r^{(0)}_{1,2} \big) =$
(0.19, 0), (0.15, 0.07), (0.04, 0.15), (0.22, 0.15), (0.30, 0.19), (0, 0.22), (0.07, 0.26), (0.11, 0.33).}
\label{fig:em_convergence}
\end{figure}

For the second type of sequence, we have provided Figure \ref{fig:em_convergence}
to confirm that the estimate obtained is the global maximum.
This figure depicts how the likelihoods increase
when the EM algorithm starts from different starting points.
As the starting point $R^{(0)}=\big( r_{k,l}^{(0)}\big) $, we chose:
(a) the estimator $\widetilde R$ in (\ref{inconsistent}) of 
Remark \ref{rem:inconsistent} (as indicated in Algorithm \ref{alg:em});
(b) $r^{(0)}_{k,l} \equiv 1/6$; and 
(c) 8 points randomly chosen from the $R$-region defined by (\ref{sumr}) 
(see the legend of Figure \ref{fig:em_convergence}).
From this figure, we can see that
the limit starting from the estimator $\widetilde R$ in (\ref{inconsistent})
attains the maximum of the likelihood function.

We also conducted a maximization of the likelihood function 
by means of a numerical grid search for $R$ with $(m,n) = (2,3)$. 
Our calculations indicate that the maximum likelihood obtained 
by the EM algorithm is the global maximum.

\subsection{Illinois state education data}
\label{subsec:illinois}

\begin{figure}[htbp]
\begin{center}
\begin{tabular}{cc}
 \includegraphics[width=0.55\linewidth]{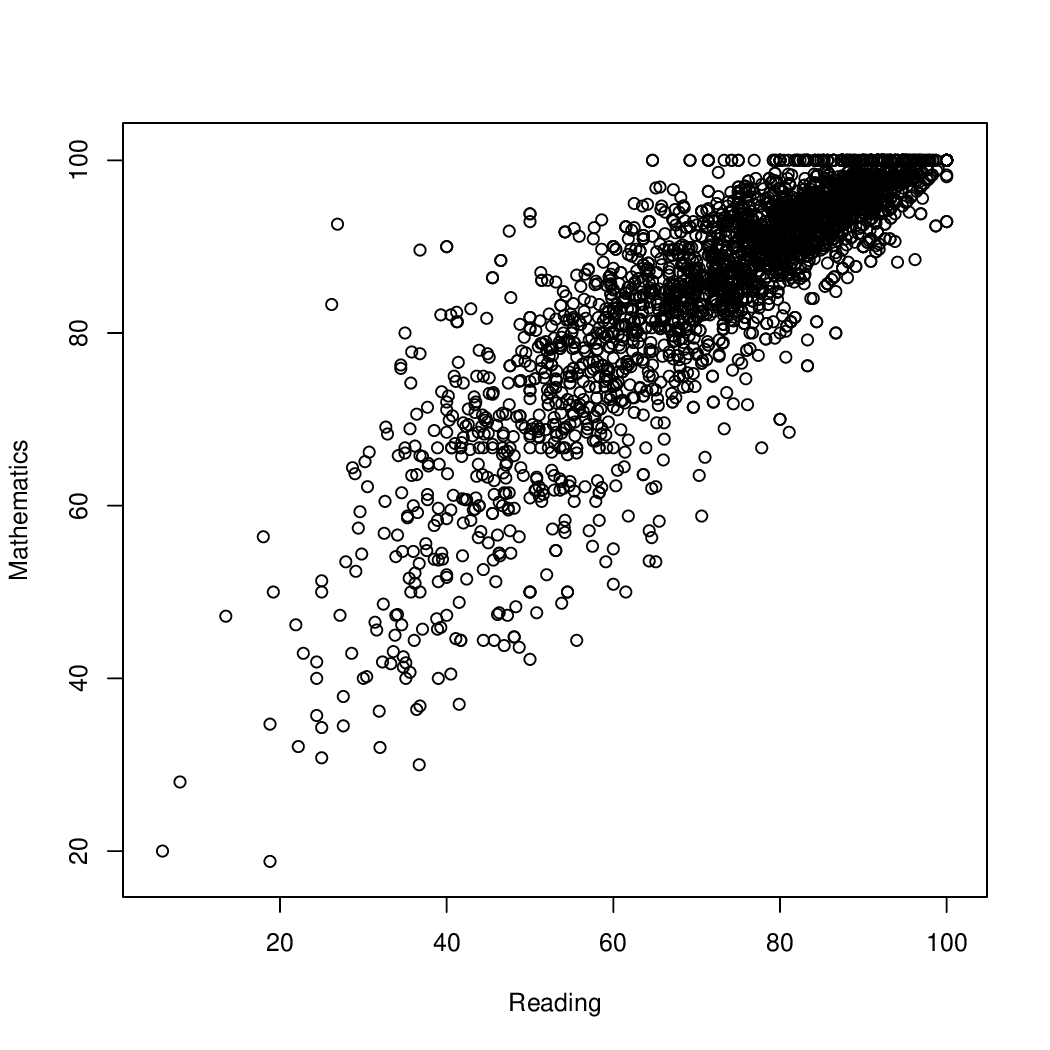} &
 \includegraphics[width=0.55\linewidth]{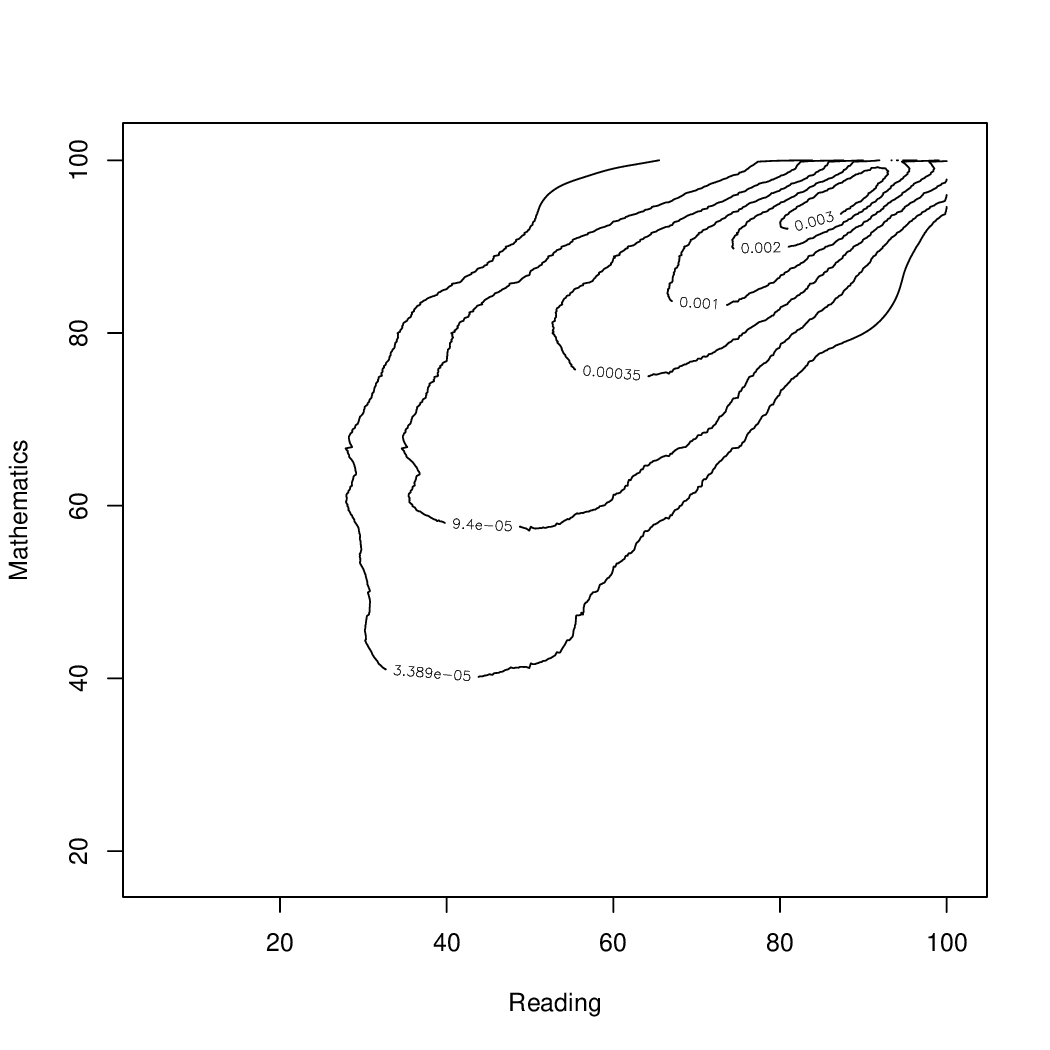}
\end{tabular}
\end{center}
\vspace{-0.3cm}
\caption{ISAT percent meeting or exceeding standards. \newline
 (Left: data plot.  Right: estimated density contour.)}
\label{fig:illinois}
\end{figure}

The second example is to estimate the joint 
density function of some Illinois Standards Achievement Test (ISAT) scores
which are available from the website of the \cite{Illinois}.
We use the ISAT performance results for reading and mathematics
in Grade 3 of $N=2991$ public schools and districts in 2009. For each school or district, 
the percentages of students meeting or exceeding test standards are tabulated
(see Table \ref{tab:isat}).
\begin{table}
\caption{2009 ISAT (Illinois Standards Achievement Test). \newline
The percentage of student scores meeting or exceeding standards in reading and mathematics, 
Grade 3 for $N=2991$ schools and districts.
}
\label{tab:isat}
\begin{center}
\begin{small}
\begin{tabular}{lccc}
\hline
District name/ School name  & Reading & Mathematics & County\\
\hline
Payson CUSD 1                & 78.6 &  88.4 & Adams \\
Seymour Elementary School    & 78.6 &  88.4 & Adams \\
Liberty CUSD 2               & 84.6 & 100.0 & Adams \\
Liberty Elementary School    & 84.6 & 100.0 & Adams \\
Central CUSD 3               & 63.6 & 87.9  & Adams \\
Central 3-4 Middle School    & 63.6 & 87.9  & Adams \\
CUSD 4                       & 69.6 & 71.4  & Adams \\
Greenfield Elementary School & 69.6 & 71.4 & Adams \\
Quincy SD 172                & 73.0 & 86.9 & Adams \\
Adams Elementary School      & 67.1 & 79.7 & Adams \\
Dewey Elementary School      & 75.4 & 93.4 & Adams \\ 
Ellington Elementary School  & 90.1 & 94.4 & Adams \\
\hfil$\vdots$\hfil & $\vdots$ & $\vdots$ & $\vdots$ \smallskip \\
\hline
\end{tabular}
\end{small}
\end{center}
\end{table}
The data are plotted in Figure \ref{fig:illinois} (left).
Each point indicates a public school or district.

We first estimate the density functions and (cumulative) distribution 
functions by the kernel method and the empirical distribution function.
Pearson's correlation and the sample rank correlation of the data are
$0.853$ and $0.851$, respectively.  Because of these high sample 
correlations, we use $H^{+}_n(x,y)$ in (\ref{Hpm}), the largest 
correlation model.  The estimated density function, 
$h^+(x,y;\widehat q,\widehat n)$, 
is plotted in Figure \ref{fig:illinois} (right).
Using the EM algorithm in Section \ref{subsec:parametric},
we obtain the estimates $(\widehat q, \widehat n) = (0.919,\,17)$.
The approximate variance of $\widehat q$ is $9.06\times 10^{-5}$.
The rank correlation under the estimated model is
$\widehat q (\widehat n-1)/(\widehat n+1) = 0.817$.

Let $\widehat q(n)$ be the MLE of $q$ under the model with $n$ fixed 
as in \cite{Baker08}.  Table \ref{tab:qn} lists the MLEs $\widehat q(n)$ 
and the corresponding log-likelihoods (profile likelihoods) for 
$2 \le n \le 20$, and we see that the log-likelihood is maximized at 
$n=16$.  
Although this is different from the estimator $\widehat n=17$ above,
the difference between the values of the likelihood at these two 
estimates is small, and this shows that Algorithm \ref{alg:qn} 
works well in this practical setting.  

\begin{table}[htbp]
\caption{MLE and profile log-likelihood when $n$ is fixed. \newline
(The maximum value is indicated with a box.)}
\label{tab:qn}
\begin{center}
\begin{small}
\begin{tabular}{ccc}
\hline
$n$ & $\widehat q(n)$ & Log-likelihood \\
\hline
2  & 1.000 & --22580.84 \\
5  & 1.000 & --21812.38 \\
10 & 0.980 & --21477.13 \\
12 & 0.970 & --21439.37 \\
14 & 0.949 & --21422.39 \\
15 & 0.939 & --21418.51 \\
16 & 0.929 & \fbox{--21416.81} \\
17 & 0.919 & --21416.88 \\
18 & 0.909 & --21418.41 \\
19 & 0.899 & --21421.13 \\
20 & 0.889 & --21424.85 \\
\hline
\end{tabular}
\end{small}
\end{center}
\end{table}

The analysis above assumes that the scores are continuous variables.
However, as shown in Table \ref{tab:isat}, the scores are rounded off
to the nearest one-tenth value.  Therefore, in practice, the variables 
are discrete, take values $k/10$, $k=0,1,\ldots,1000$, and there are 
many ties in this data set.  Also, 
the number of unique values for $x_i$, $y_i$, and
$(x_i,y_i)$ are $602$, $456$ and $2260$, respectively, among $N=2991$
schools and districts.
Applying the EM algorithm for the discrete case, i.e., 
Algorithm \ref{alg:qn} 
with (\ref{E-continuous}) replaced by (\ref{E-discrete}), we obtain
the estimates $(\widehat q,\widehat n)=(0.933,\,18)$.
The rank correlation under the discrete model is $0.835$, which is slightly
closer to the sample rank correlation than the one under the continuous model.
In both cases, we see that
the estimated rank correlations are slightly lower
than the sample rank correlation of the data.
However, considering the few number of parameters in the models, 
the differences may be acceptable.

\begin{figure}[htbp]
\begin{center}
\begin{tabular}{cc}
 \includegraphics[width=0.55\linewidth]{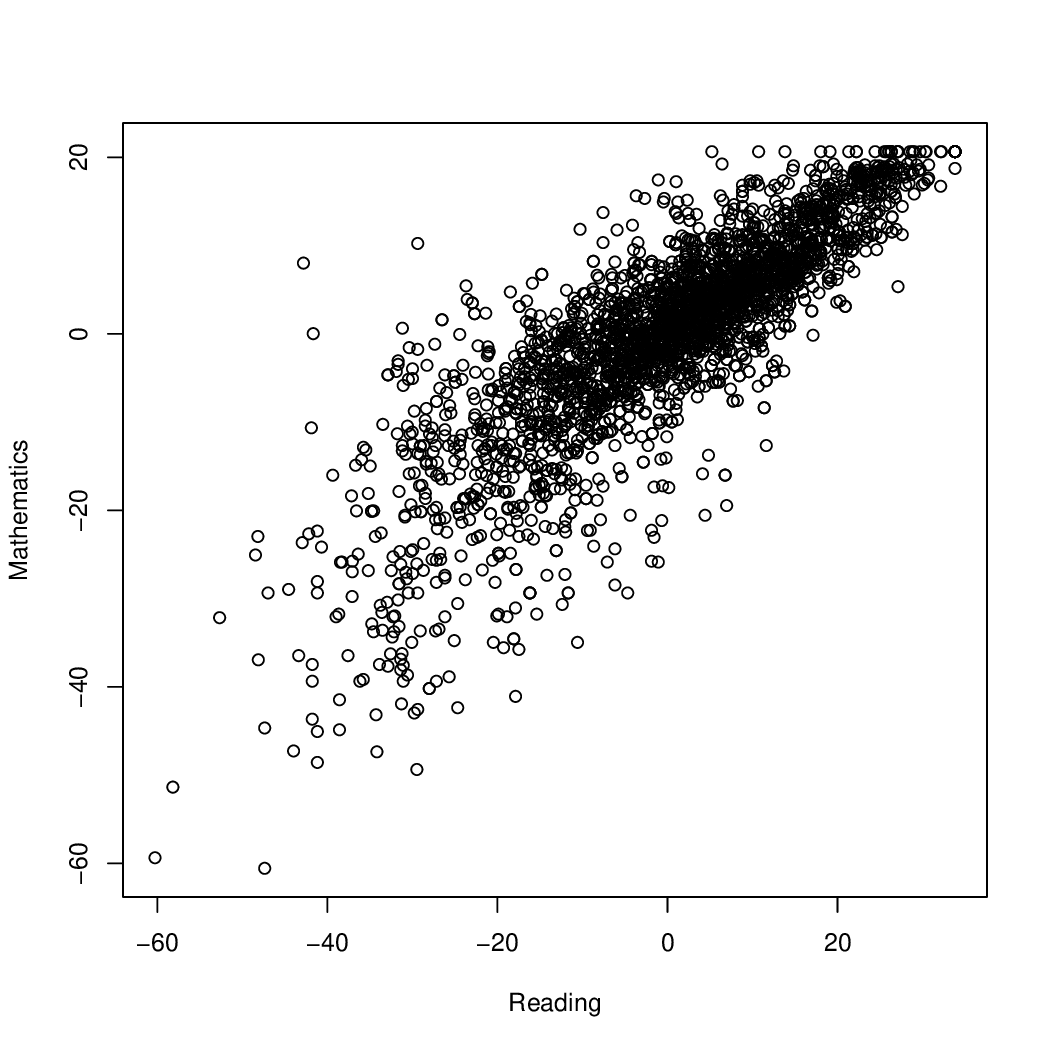} &
 \includegraphics[width=0.55\linewidth]{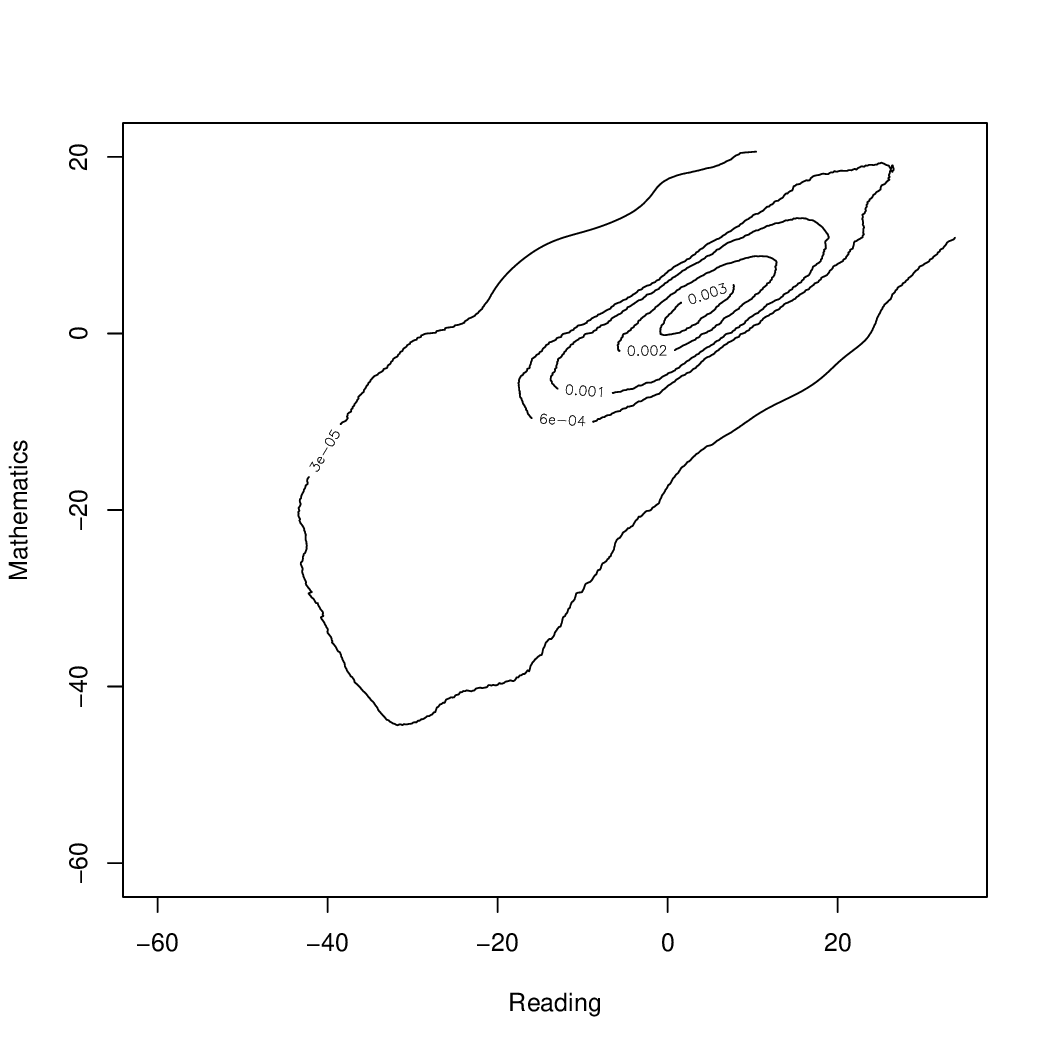}
\end{tabular}
\end{center}
\vspace{-0.3cm}
\caption{
Residuals of ISAT percent meeting or exceeding standards. \newline
 (Left: residuals by the simple regressions.  Right: estimated density contour.)}
\label{fig:illinois_county}
\end{figure}

The ISAT data set also contains the names of 102 counties to which the 2991 schools
and districts belong.
Since the high positive correlation may be caused by county effect,
we analyze residuals obtained by simple regressions for each marginal 
using the name of the county as a covariate.
The residuals are plotted in the left panel of Figure \ref{fig:illinois_county}.

Similarly to the original data, the Pearson correlation and Spearman's rank correlation
of the residuals are calculated as $0.835$ and $0.840$, respectively.
Using the EM algorithm in Section \ref{subsec:parametric}, 
we obtain the estimates $(\widehat q,\widehat n)=(0.917,\,19)$
and the joint density contour plot shown in the right panel of Figure \ref{fig:illinois_county}.
The approximate variance of $\widehat q$ is $0.0001$. 
The rank correlation under the estimated model is $0.825$.
Baker's method of maximizing profile likelihoods
gives similar results: $(\widehat q,\widehat n)=(0.919,\,19)$.

It is reasonable that the correlation of residuals is still high,
because the coefficients of determination of the simple regressions
are not large ($0.170$ and $0.187$ for Reading and Mathematics, respectively).
Note that the correlation of estimated county effects
for Reading and Mathematics is moderate ($0.732$).

\subsection{Simulated trivariate data with interaction}
\label{subsec:trivariate}

The third
example is an artificial trivariate continuous data.
The data are generated by the following two steps.
First, we generate data $(u_{1,i},u_{2,i},u_{3,i})$, $i=1,\ldots,N$, from a 
trivariate Baker's distribution 
with the copula density
\begin{align}
\label{c3}
 c(u_1,u_2,u_3)
 = n_1 n_2 n_3 \sum_{k_1=1}^{n_1}\sum_{k_2=1}^{n_2}\sum_{k_3=1}^{n_3}
r_{k_1, k_2, k_3} \prod_{j=1}^3 b_{k_j -1, n_j -1}(u_j).
\end{align}
Here, the parameter $R=(r_{k_1, k_2, k_3})$ is defined as
\[
 r_{k_1, k_2, 1}=\frac{1}{2 n_1 n_2} \ \ (\mbox{for all $k_1,k_2$}), \qquad
 r_{k_1, k_2, 2}=\begin{cases}
 \frac{1}{2 n_1} & (\mbox{if $k_1=k_2$}), \\
 0 & (\mbox{if $k_1\ne k_2$}), \end{cases}
\]
with $n_1=n_2=20$ and $n_3=2$.
The sample size is chosen to be $N=2000$.
Also, we convert the uniform marginals to normal marginals
by the transformation 
$x_i=\Phi^{-1} (u_{1,i})$, $y_i=\Phi^{-1} (u_{2,i})$, $z_i=\Phi^{-1} (u_{3,i})$,
where $\Phi^{-1} (\cdot)$ is the quantile
function of the standard normal distribution.
We then obtain random data $(x_i,y_i,z_i)$
whose marginals have standard normal distributions.

\begin{figure}[htbp]
\begin{center}
\begin{tabular}{ccc}
\includegraphics[width=0.35\linewidth]{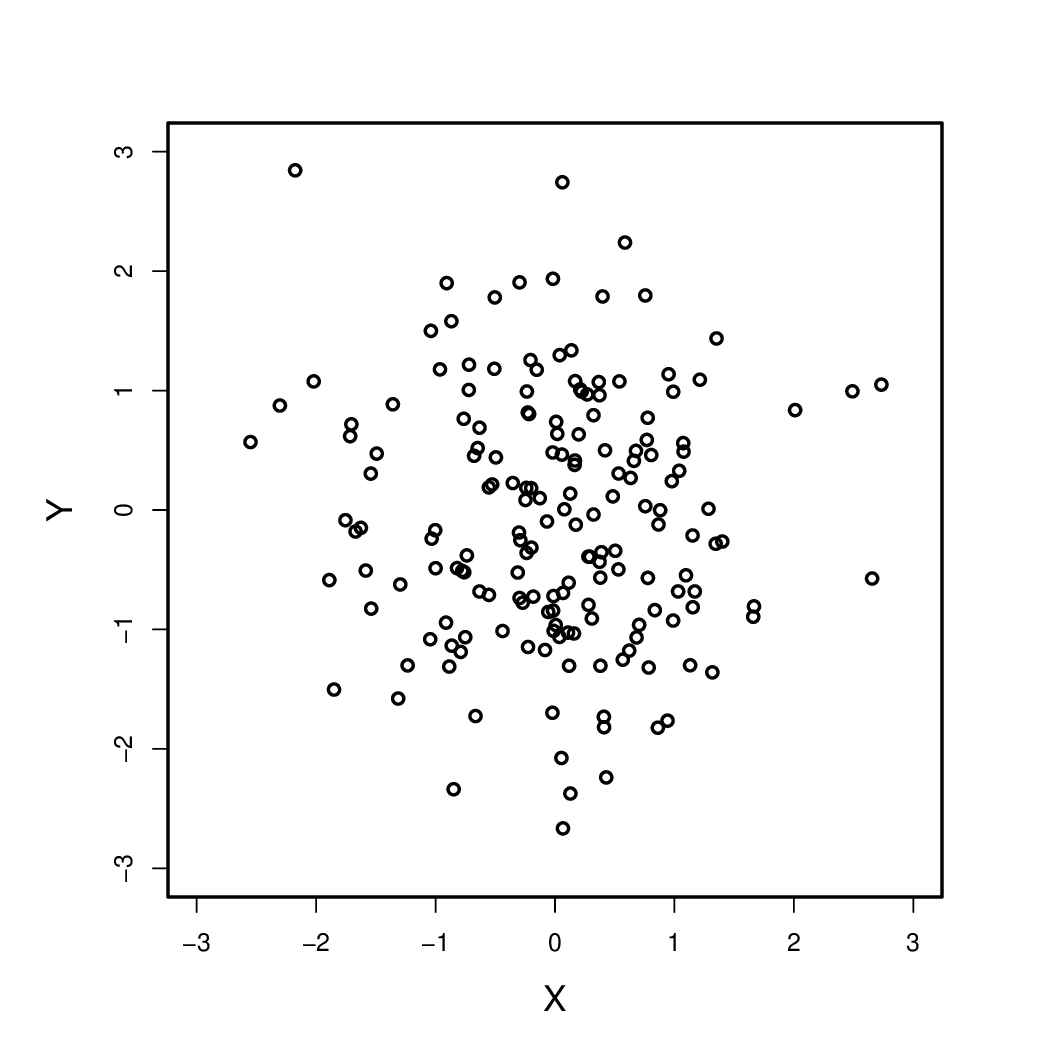} &
\includegraphics[width=0.35\linewidth]{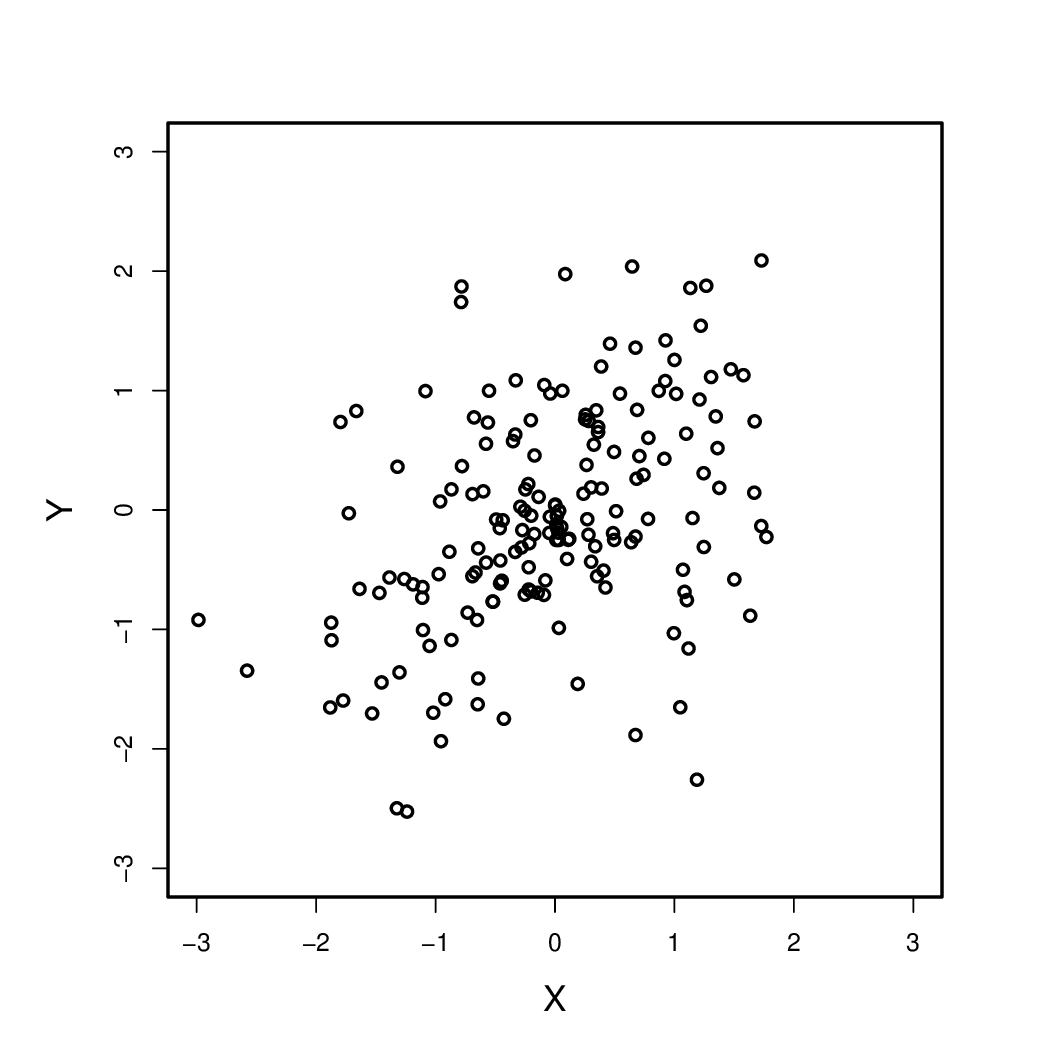} &
\includegraphics[width=0.35\linewidth]{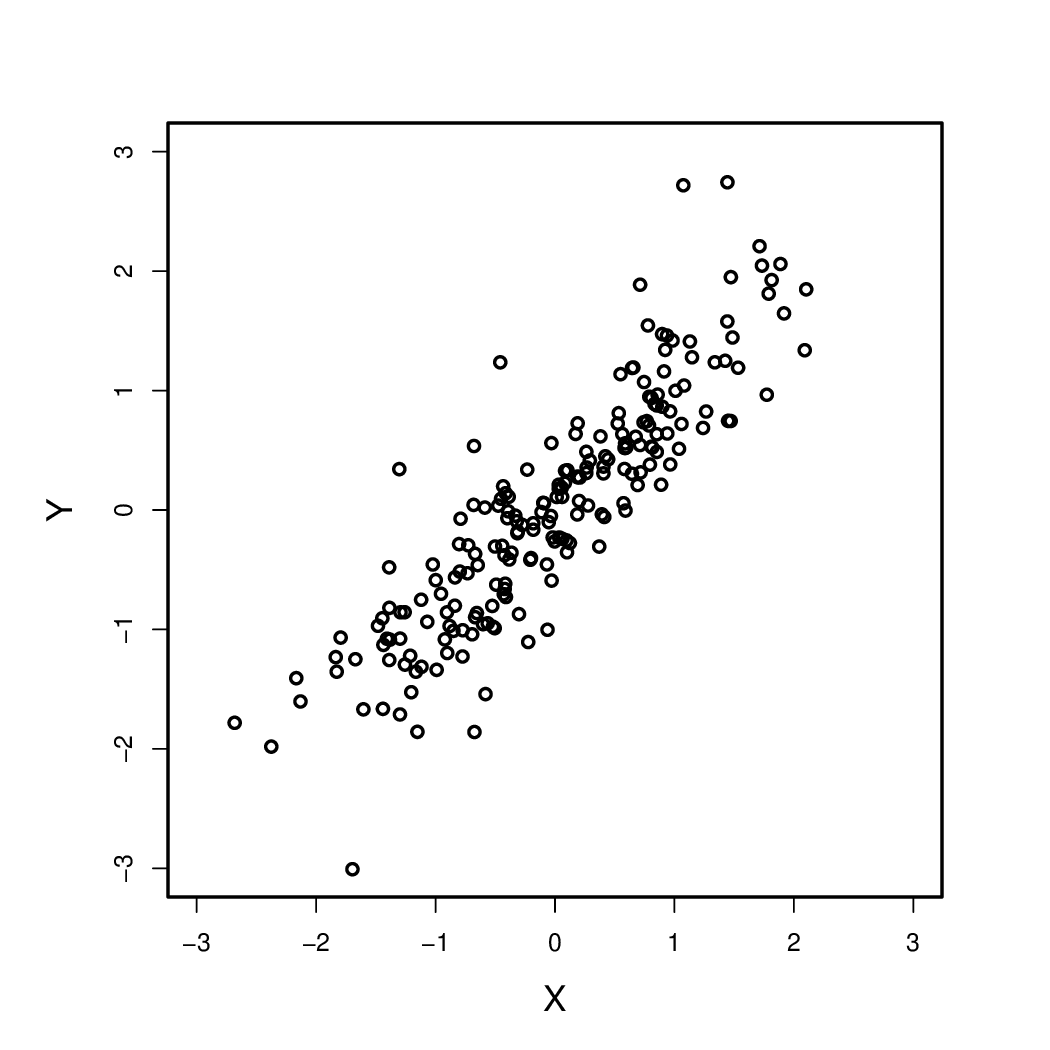} \\
\includegraphics[width=0.35\linewidth]{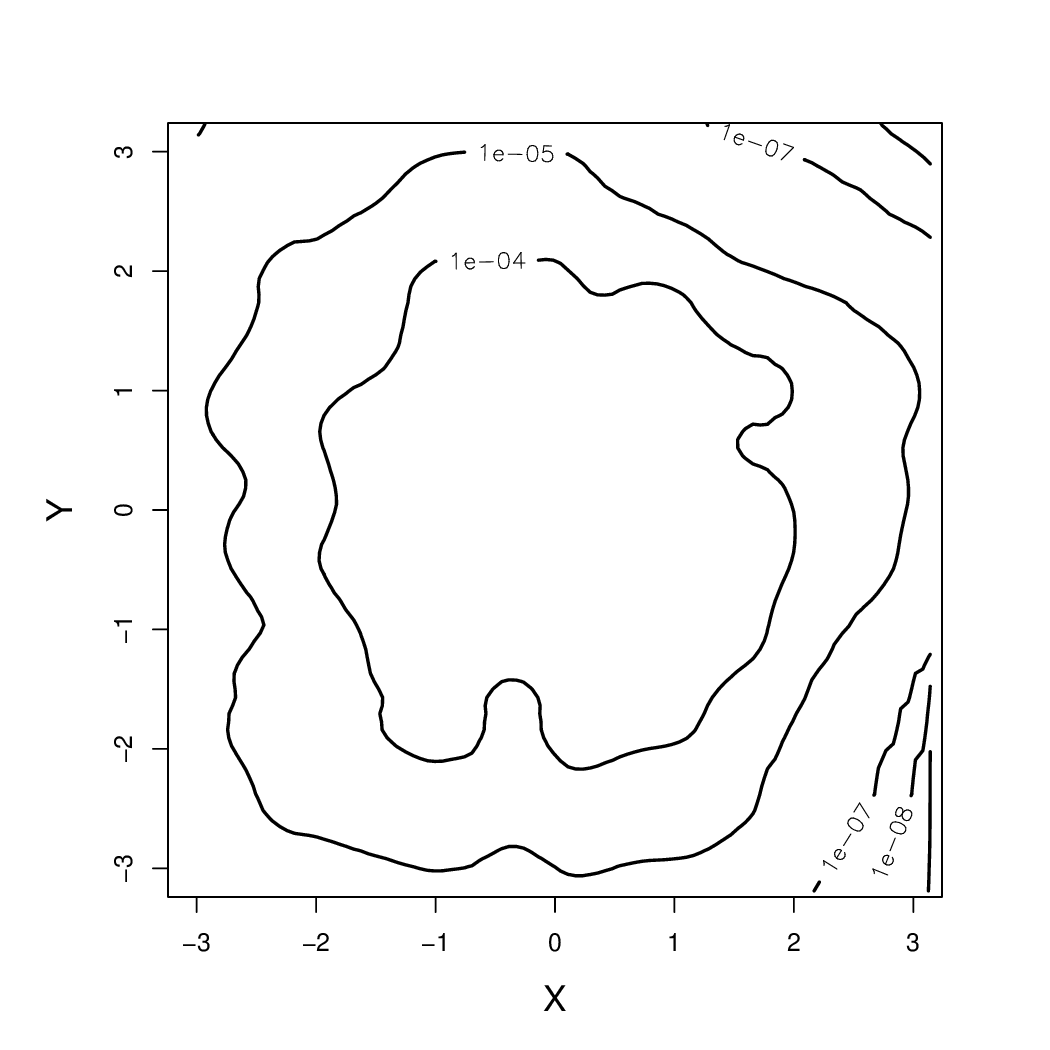} &
\includegraphics[width=0.35\linewidth]{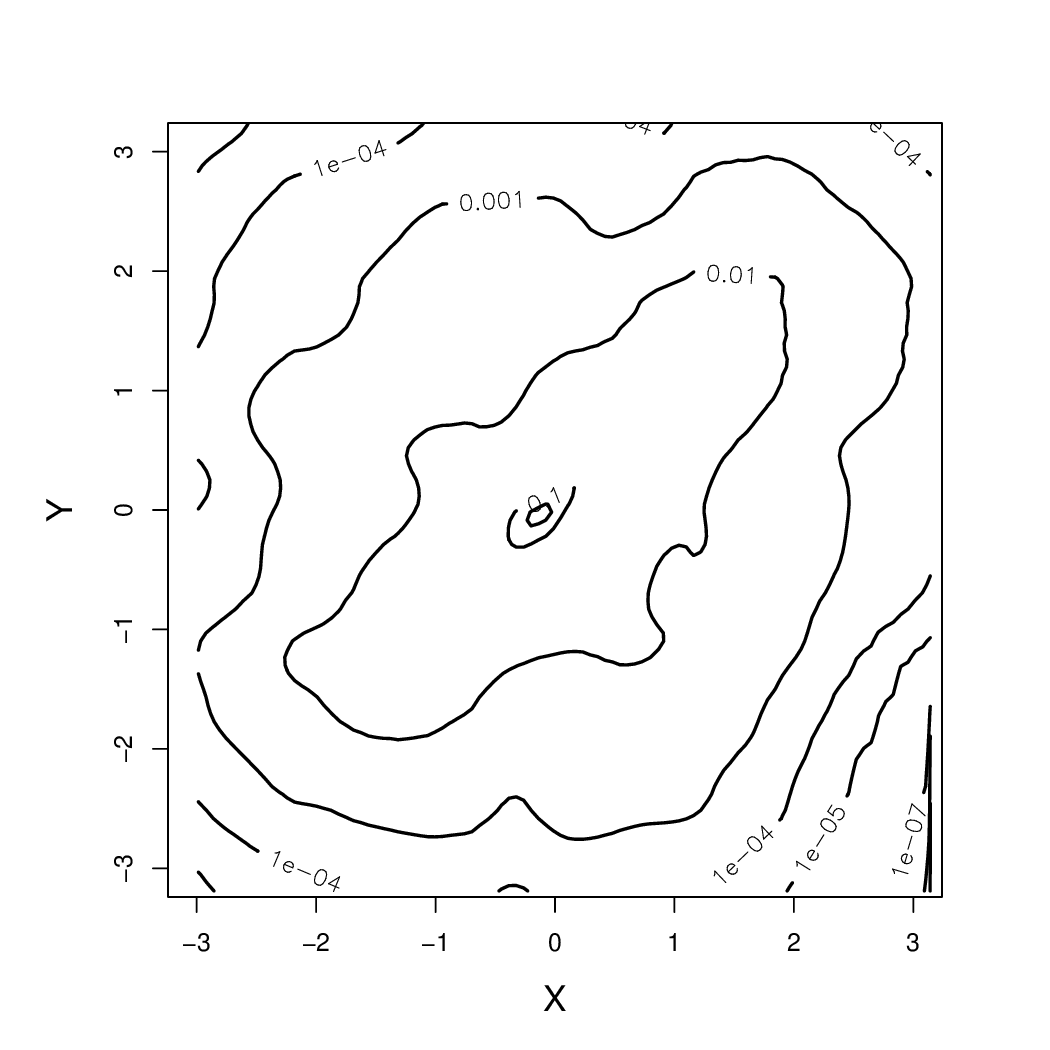} &
\includegraphics[width=0.35\linewidth]{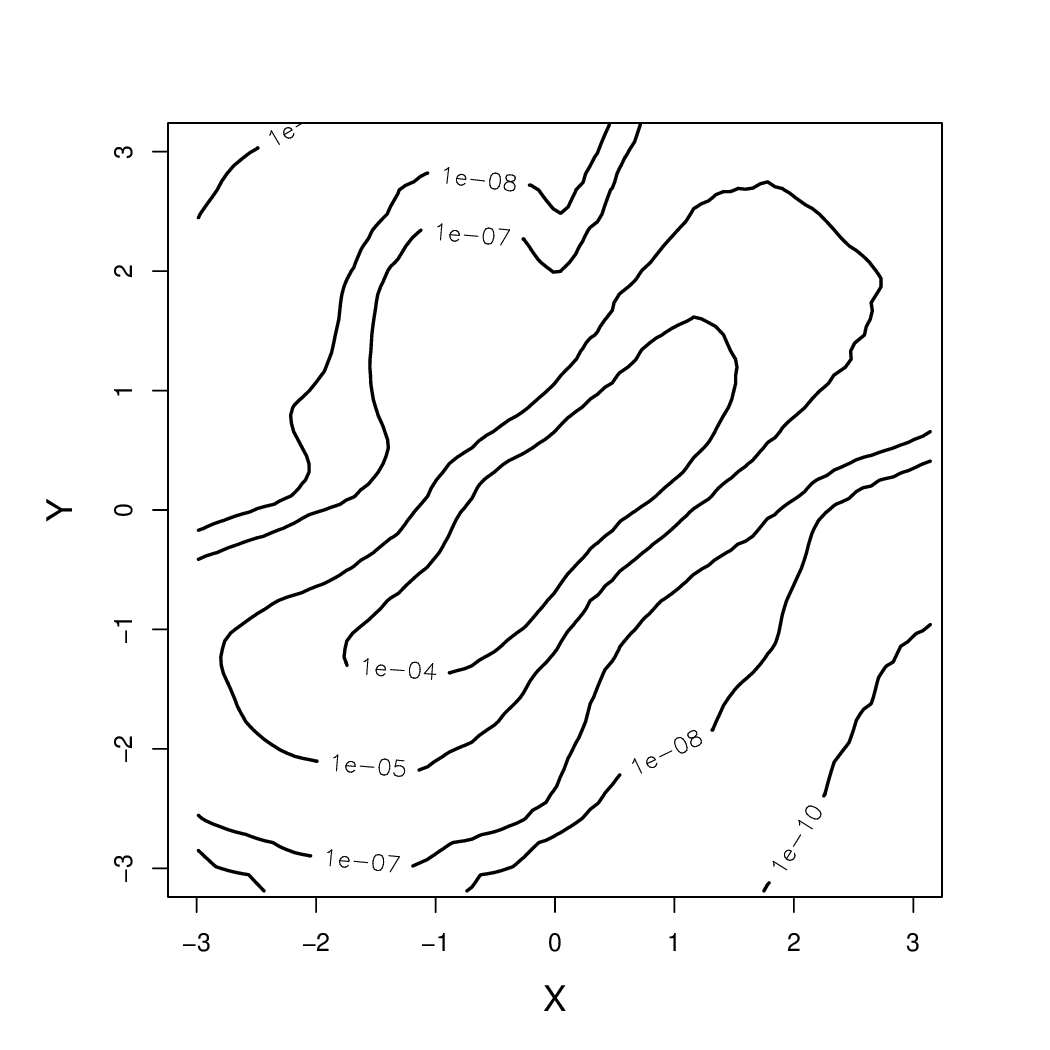} \\
\includegraphics[width=0.35\linewidth]{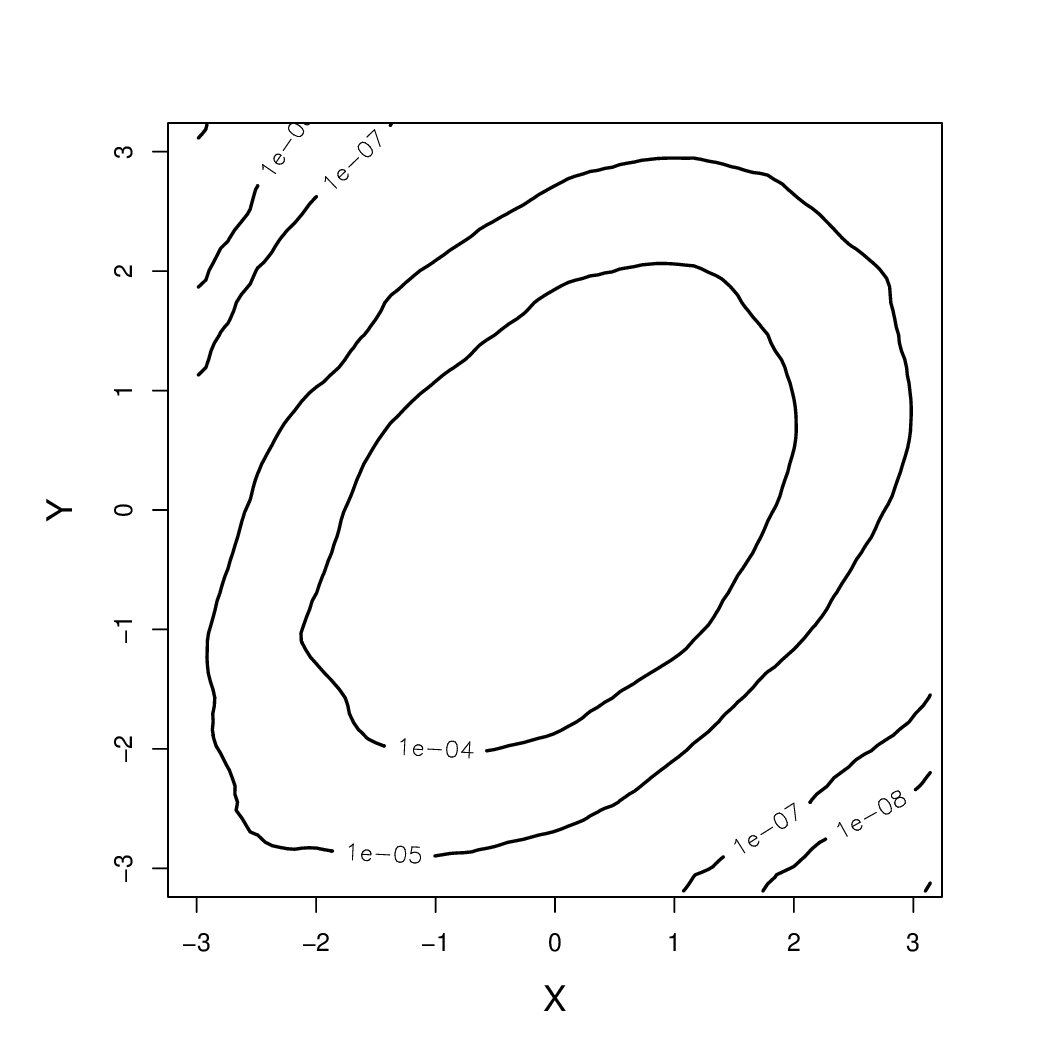} &
\includegraphics[width=0.35\linewidth]{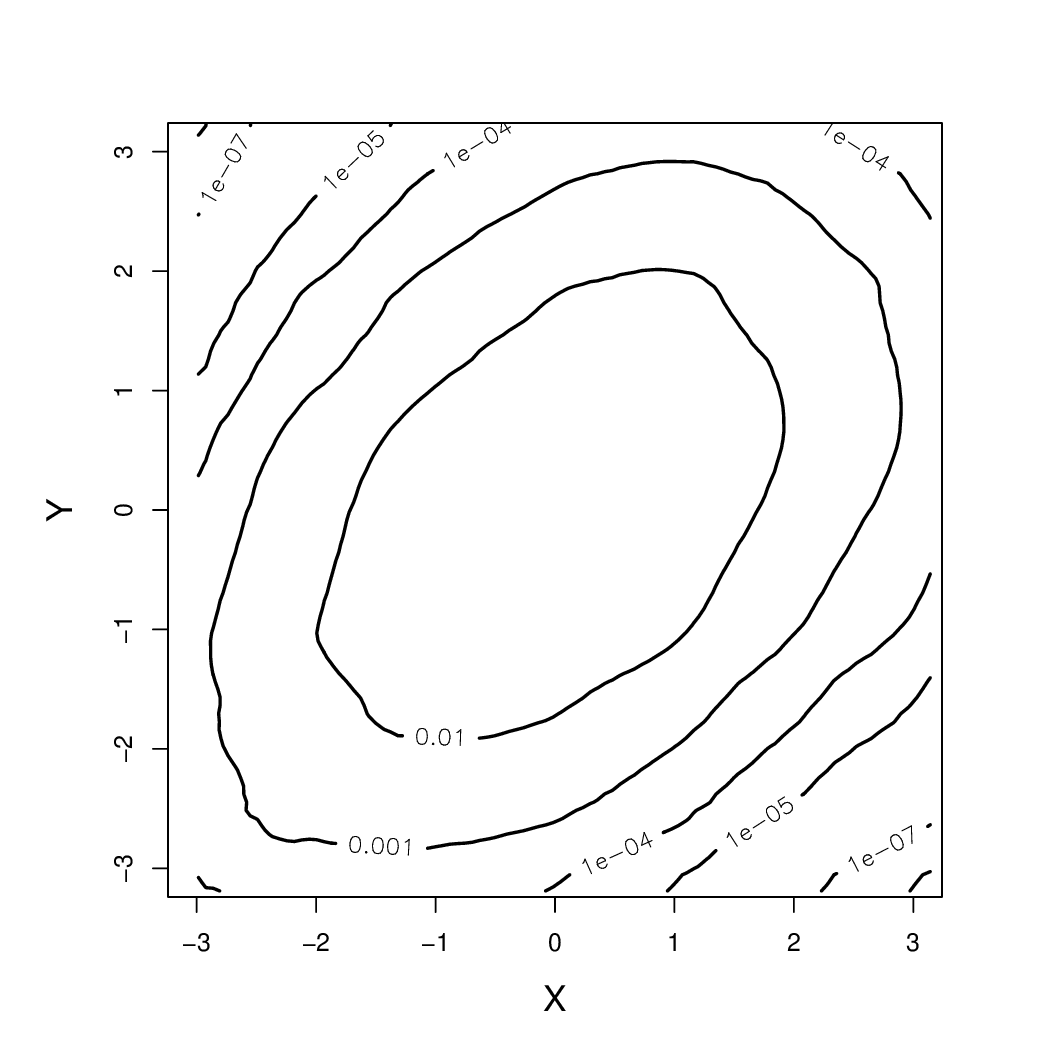} &
\includegraphics[width=0.35\linewidth]{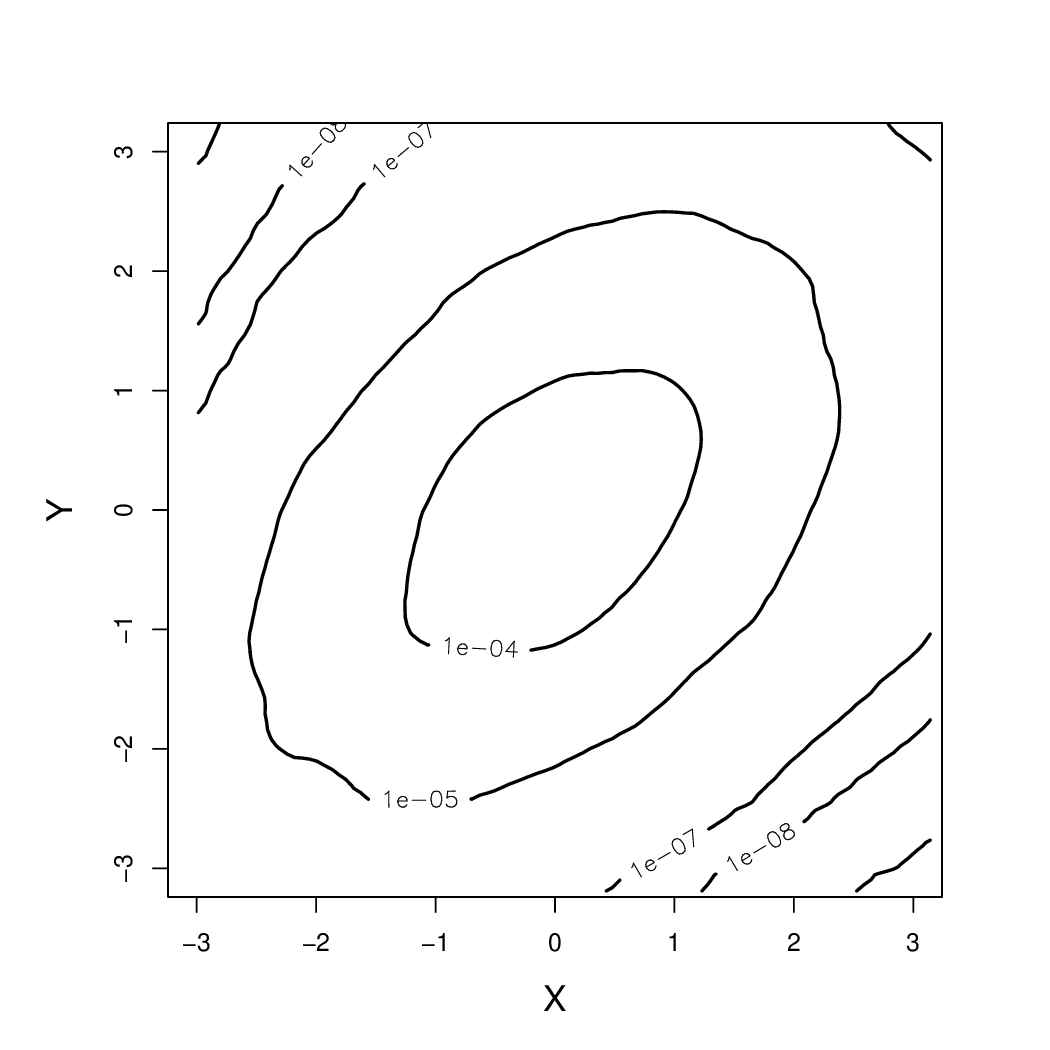}
\end{tabular}
\end{center}
\vspace{-0.3cm}
\caption{
Simulated trivariate data and estimated contours.
\newline
Scatter plots for stratified data (first row), the estimated density with a Bernstein copula (second row), and the estimated density with a Gaussian copula (third row) for three cases of $(X,Y|Z)$. \newline
Left: $Z$ is small ($Z < 0.1$); Center: $Z$ is moderately-sized ($0.45 < Z < 0.55$); Right: $Z$ is large ($Z > 0.9$).}
\label{fig:trivariate}
\end{figure}

The first row of Figure \ref{fig:trivariate} depicts scatter plots
for the first and second variates $(X,Y)$ 
stratified with the third variable $Z$.
The correlation between $X$ and $Y$ is designed to be increasing in $Z$, 
and the marginals of $(X,Z)$ and $(Y,Z)$ are independent.
From the three panels, we can see that $X$ and $Y$ are
almost independent when $Z$ is small and they are highly correlated 
when $Z$ is large.

We fit the Bernstein copula density (\ref{c3})
with an extended version of Algorithm \ref{alg:em} as well as
Algorithm \ref{alg:m} for this 3-dimensional data set.
The contours of the estimated density function are shown
in the second row of Figure \ref{fig:trivariate}, and we
see that the Bernstein copula represents well the characteristic 
of the changing correlation.
For comparison, we also plot the contours estimated with the 
Gaussian copula in the last row of the figure even though 
a Gaussian copula obviously cannot adapt to the change of correlation (i.e., the 3-way interaction).
Consequently, this example demonstrates the flexibility of the Bernstein copula
and the usefulness of the EM algorithm for 3-dimensional data.

\subsection{Trivariate uranium data}
\label{subsec:uranium}

In the fourth example, we consider the uranium data set 
which is analyzed in \cite{Cook-Johnson86}
and available from the R package \texttt{copula}.
The data set consists of concentrations of 7 elements measured from 655 water samples collected from the Montrose Quadrangle of Western Colorado.
We select three elements cobalt (Co), scandium (Sc), and cesium (Cs) as 
variables and fit the trivariate Bernstein copula model to the data. 

\begin{figure}[htbp]
\begin{center}
\begin{tabular}{ccc}
\includegraphics[width=0.35\linewidth]{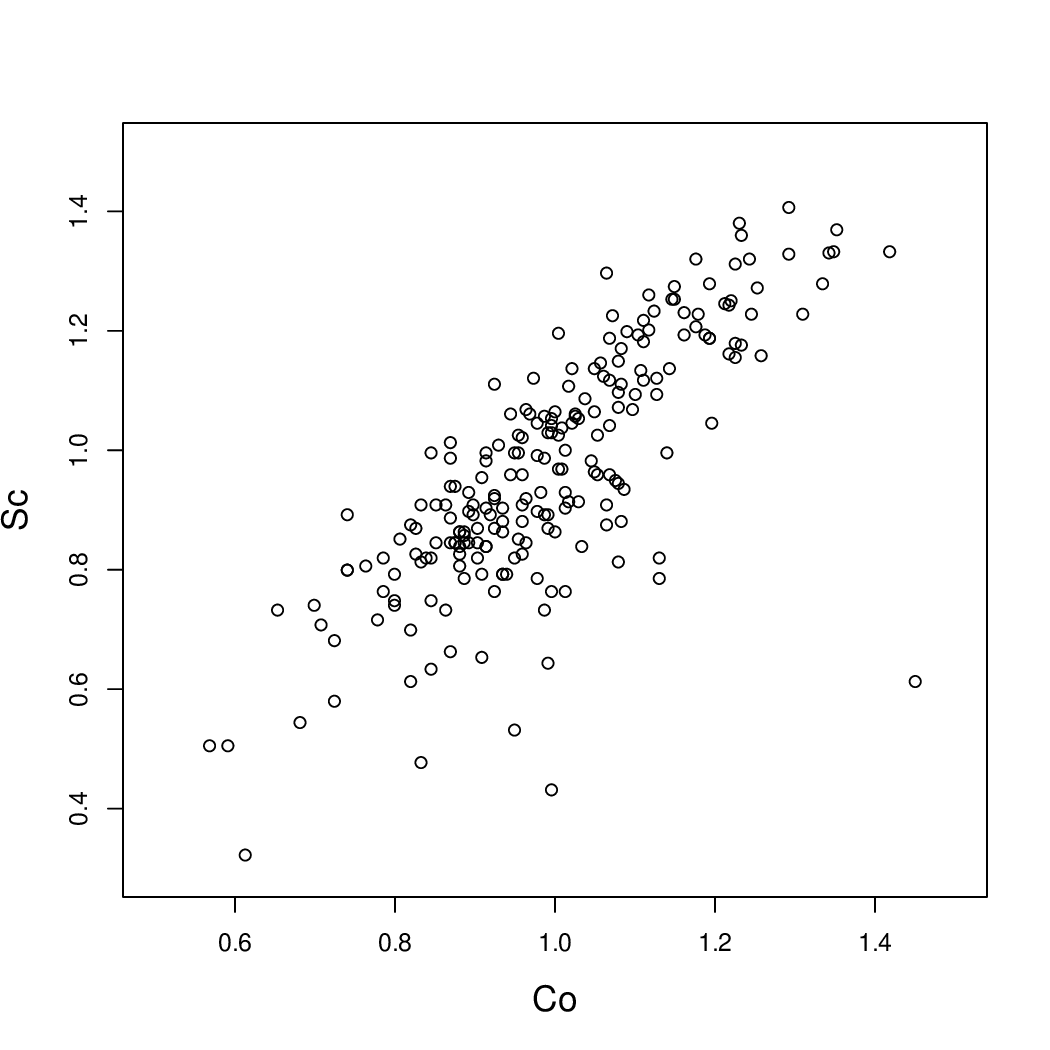} &
\includegraphics[width=0.35\linewidth]{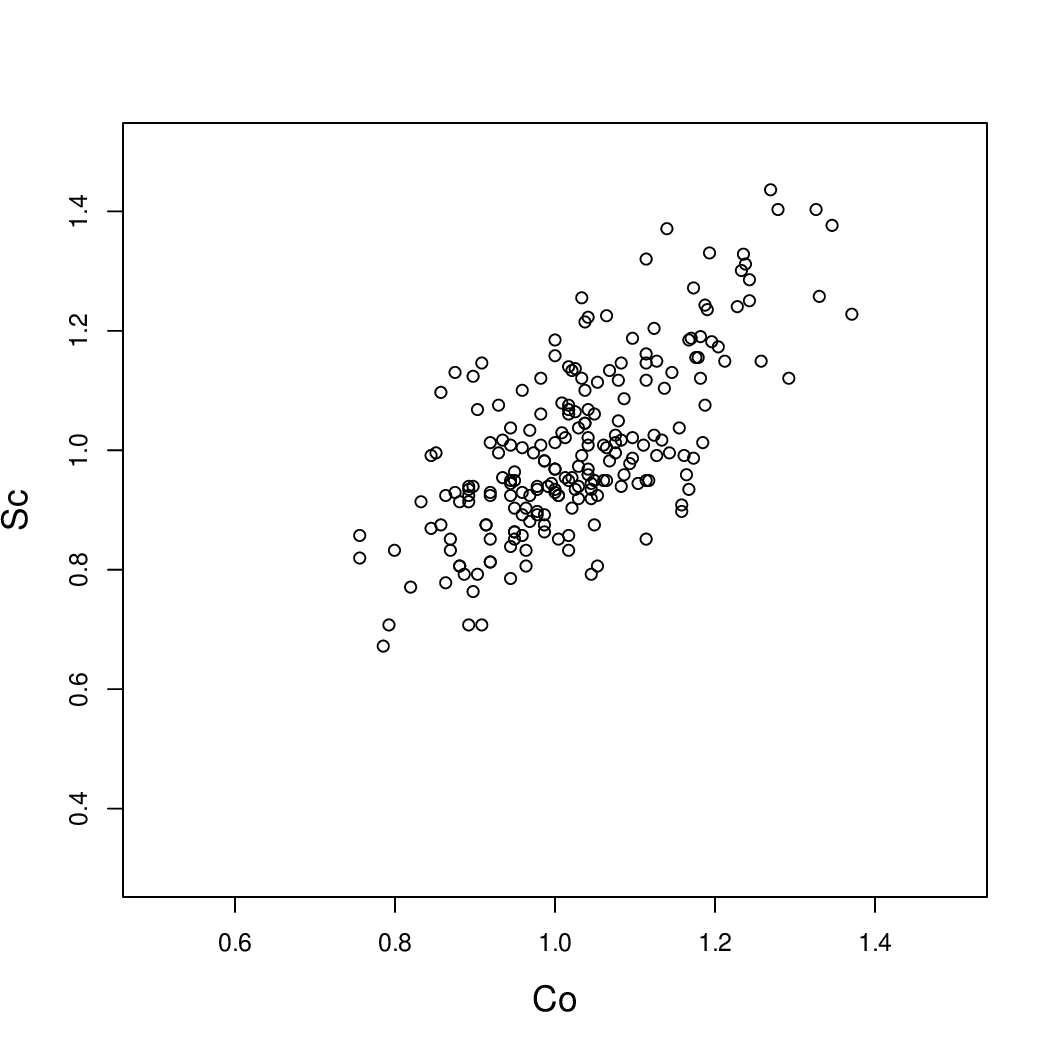} &
\includegraphics[width=0.35\linewidth]{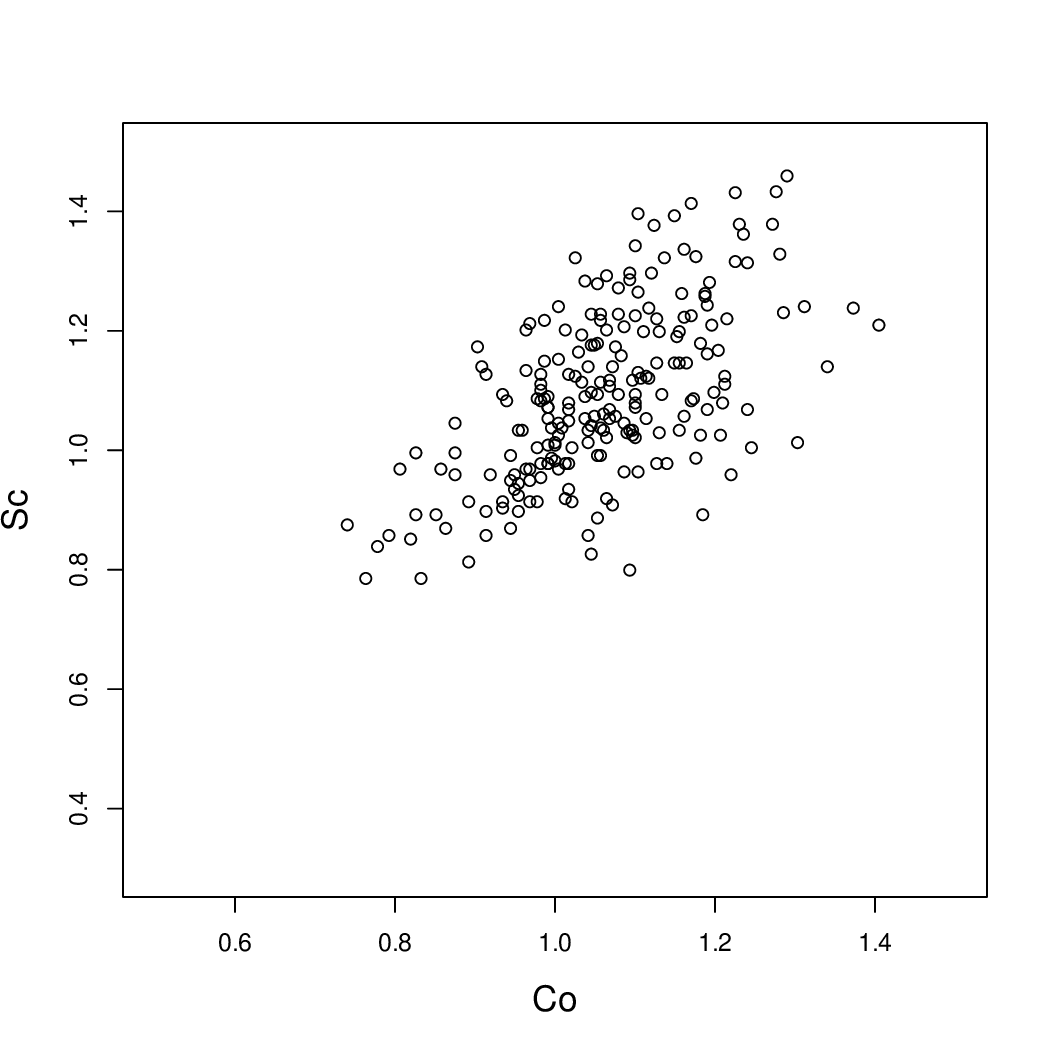} \\
\includegraphics[width=0.35\linewidth]{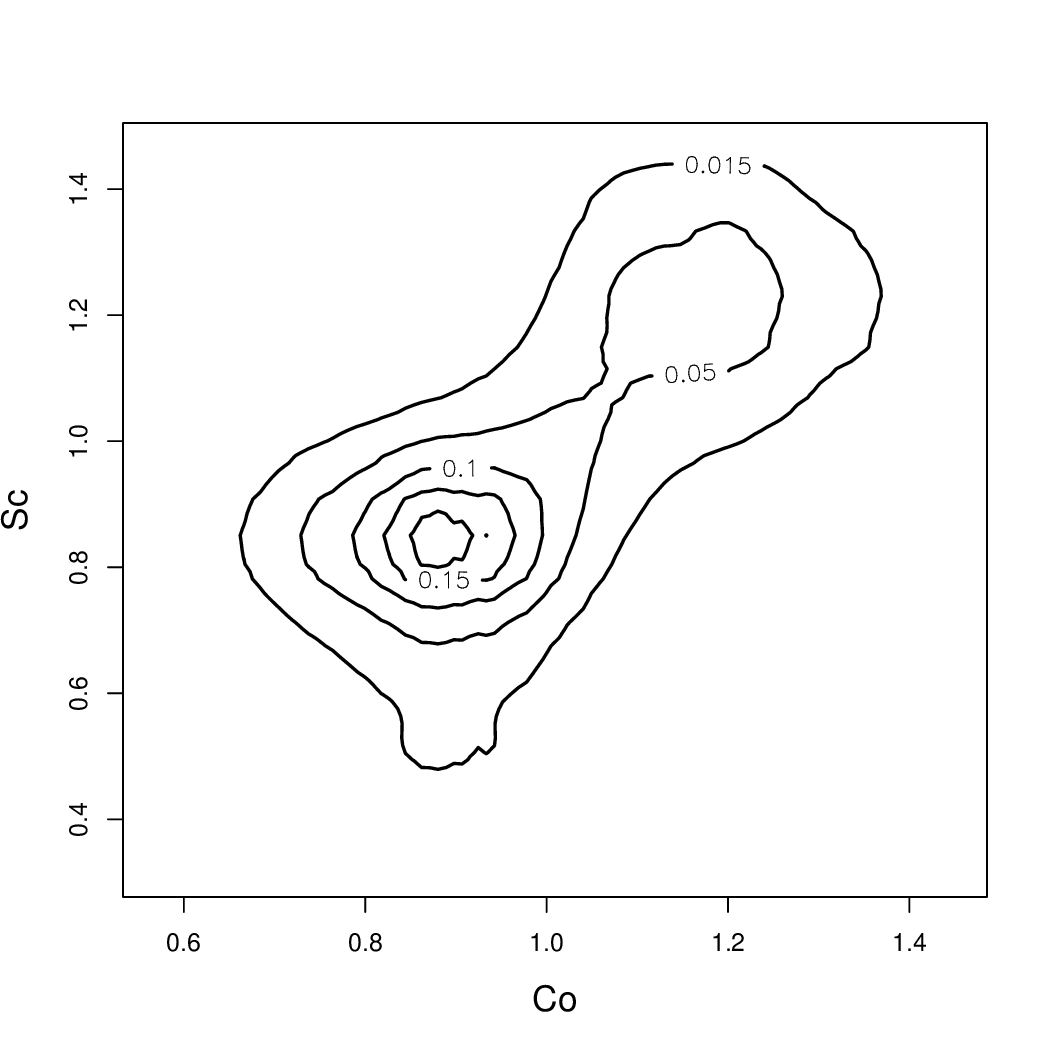} &
\includegraphics[width=0.35\linewidth]{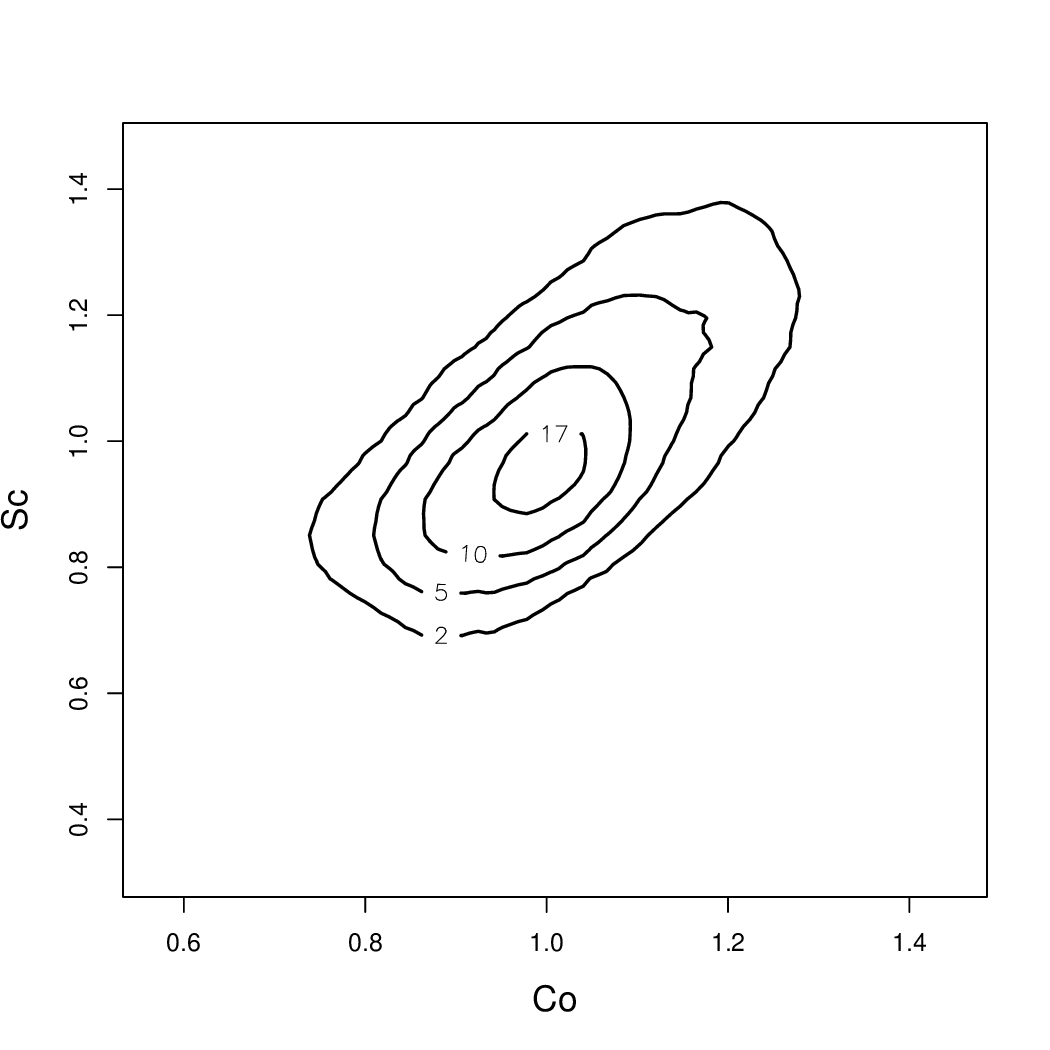} &
\includegraphics[width=0.35\linewidth]{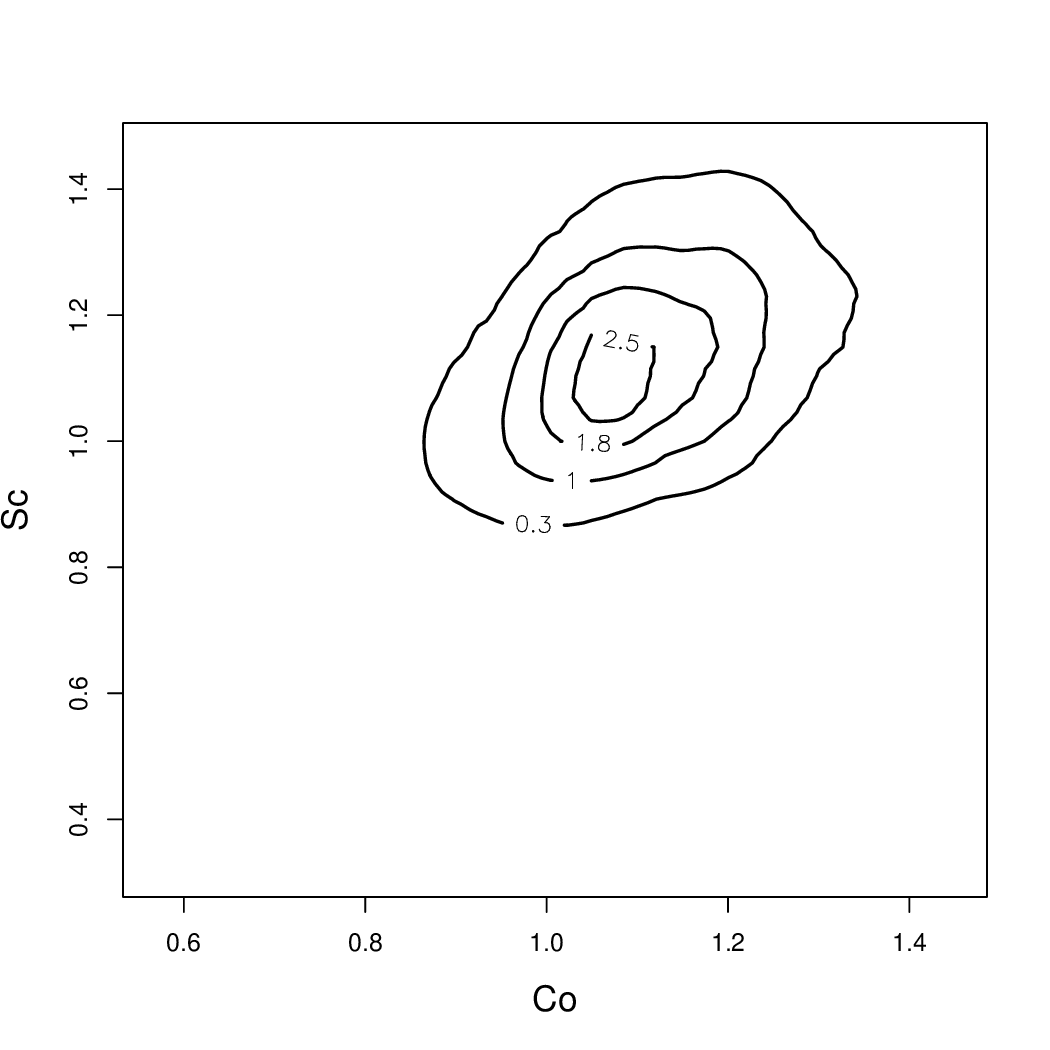} \\
\includegraphics[width=0.35\linewidth]{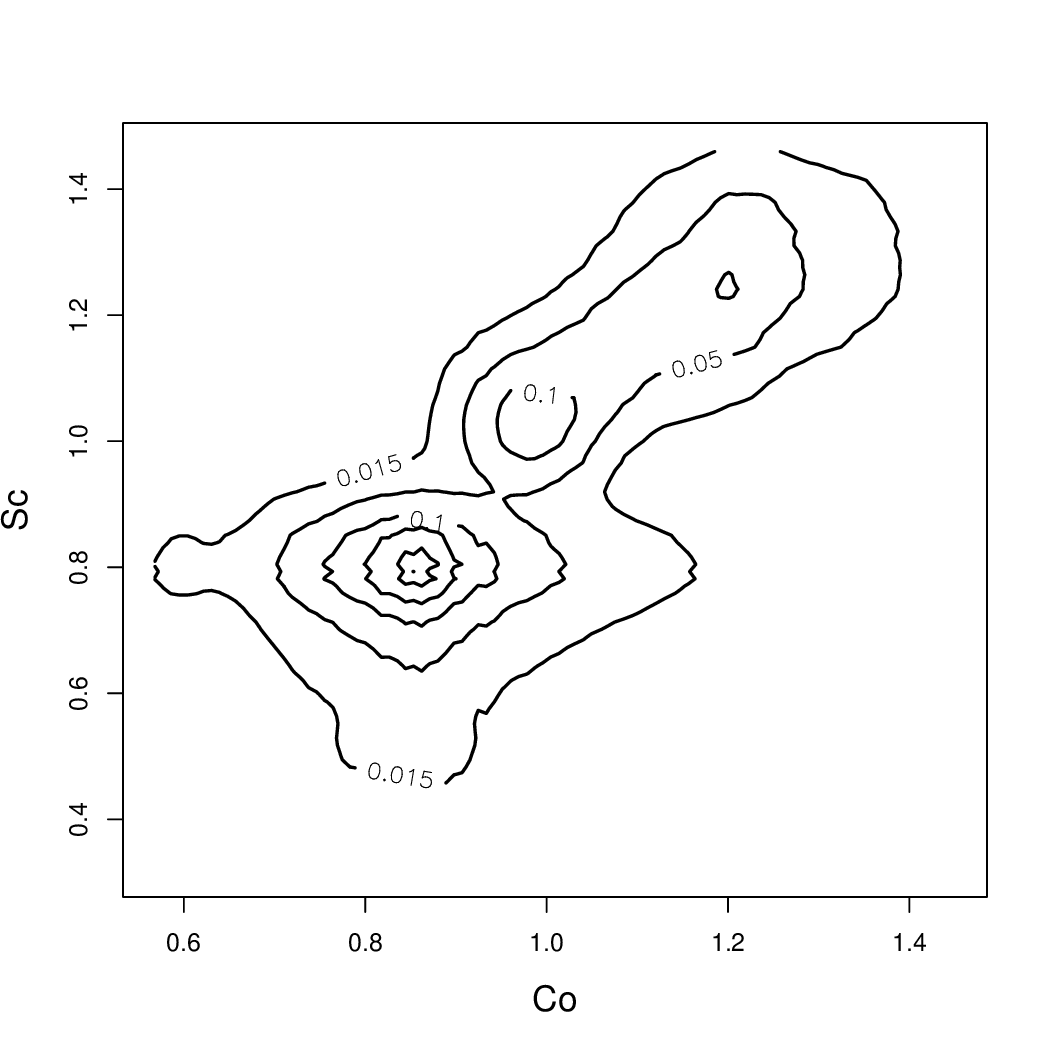} &
\includegraphics[width=0.35\linewidth]{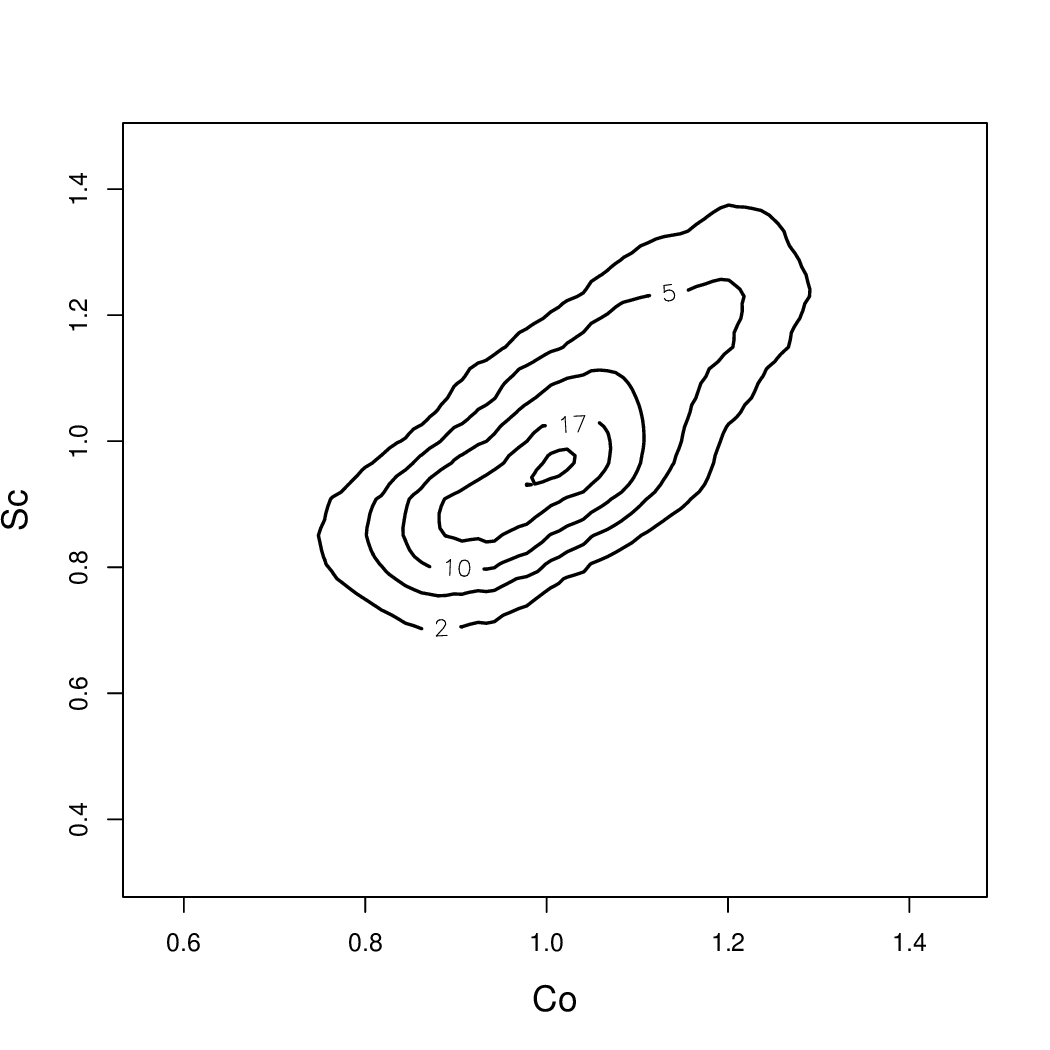} &
\includegraphics[width=0.35\linewidth]{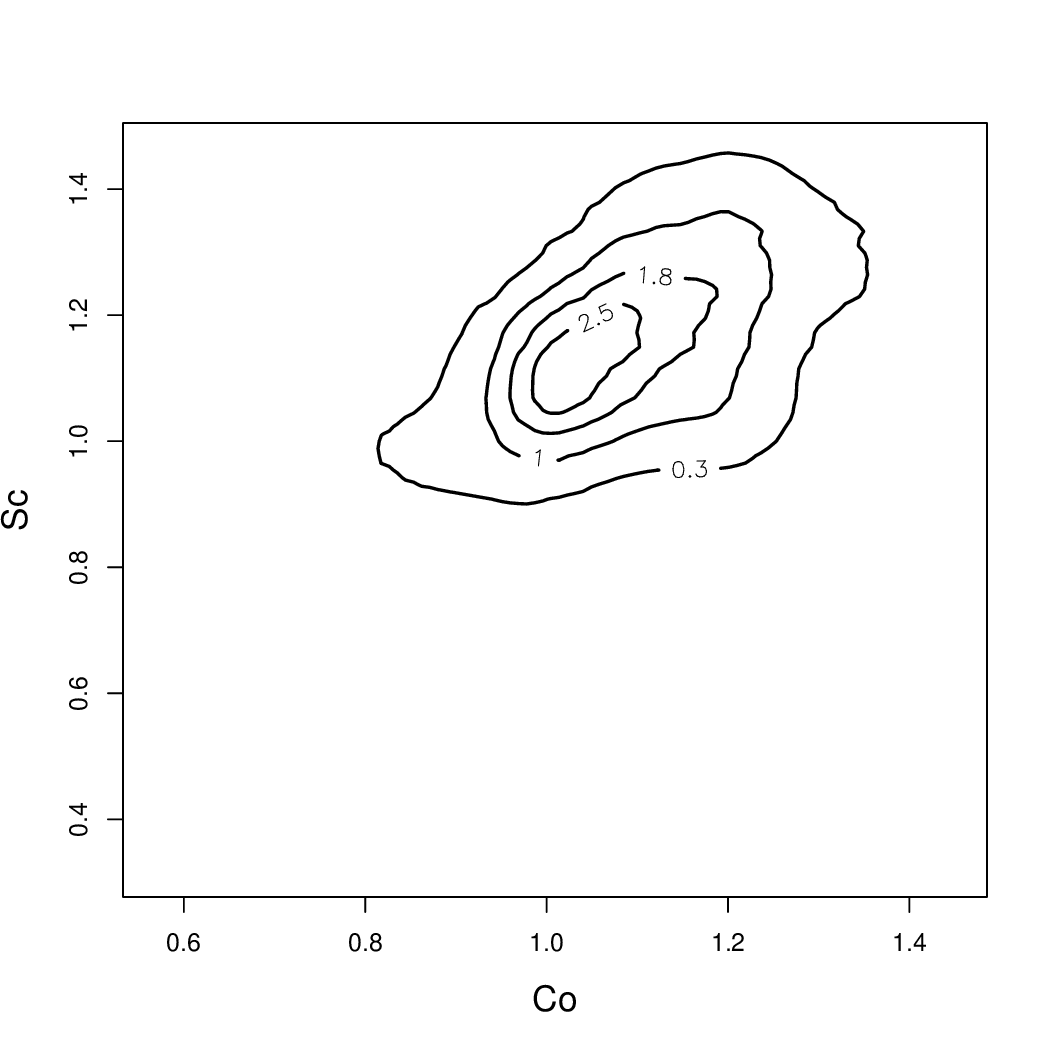} \\
\includegraphics[width=0.35\linewidth]{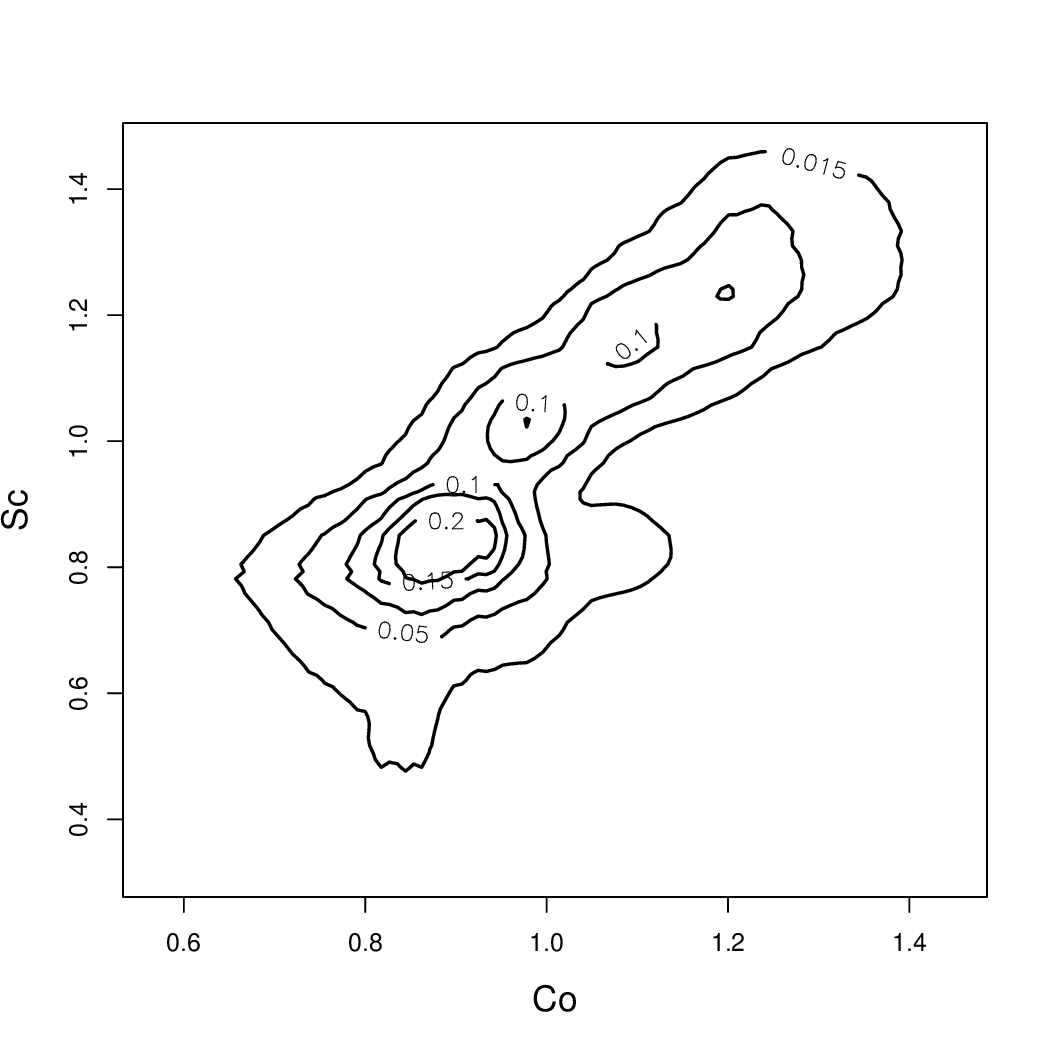} &
\includegraphics[width=0.35\linewidth]{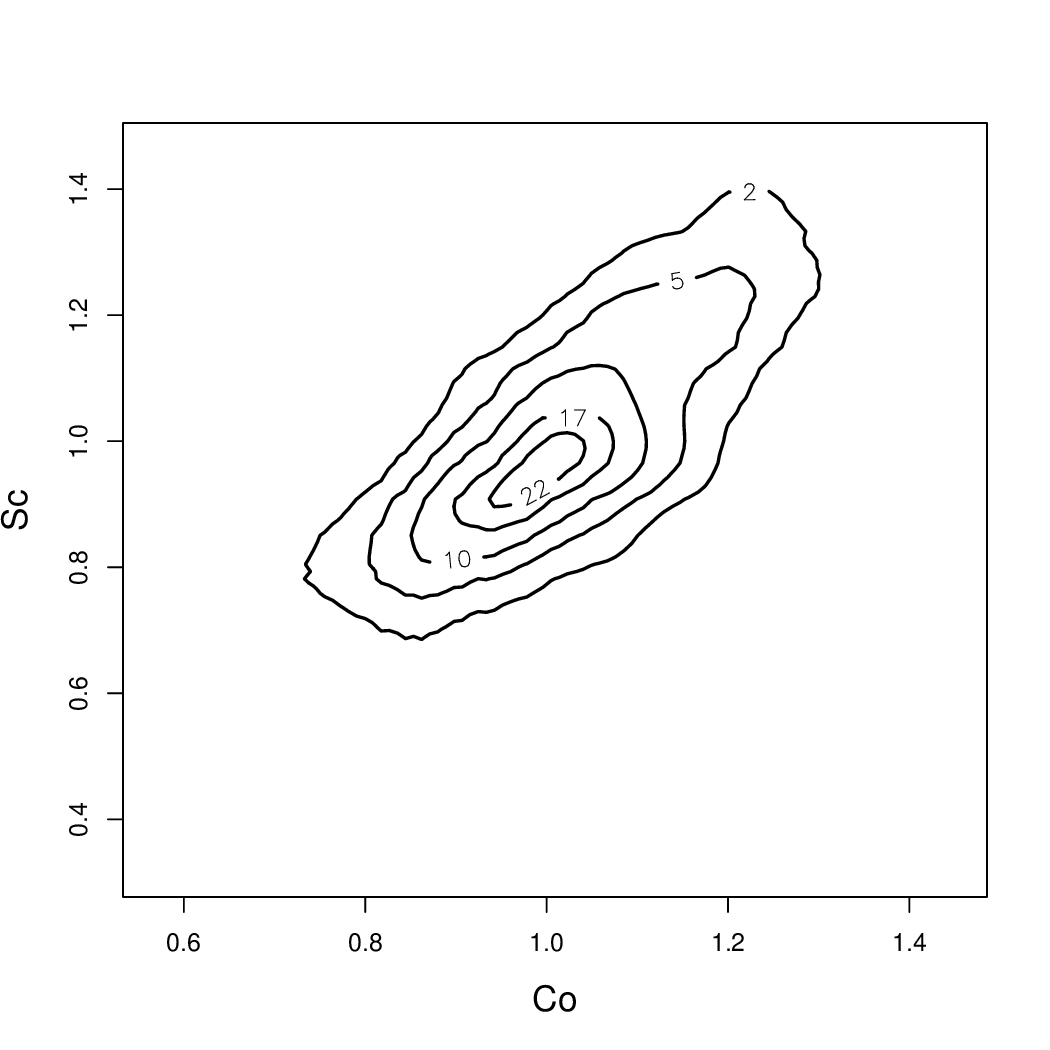} &
\includegraphics[width=0.35\linewidth]{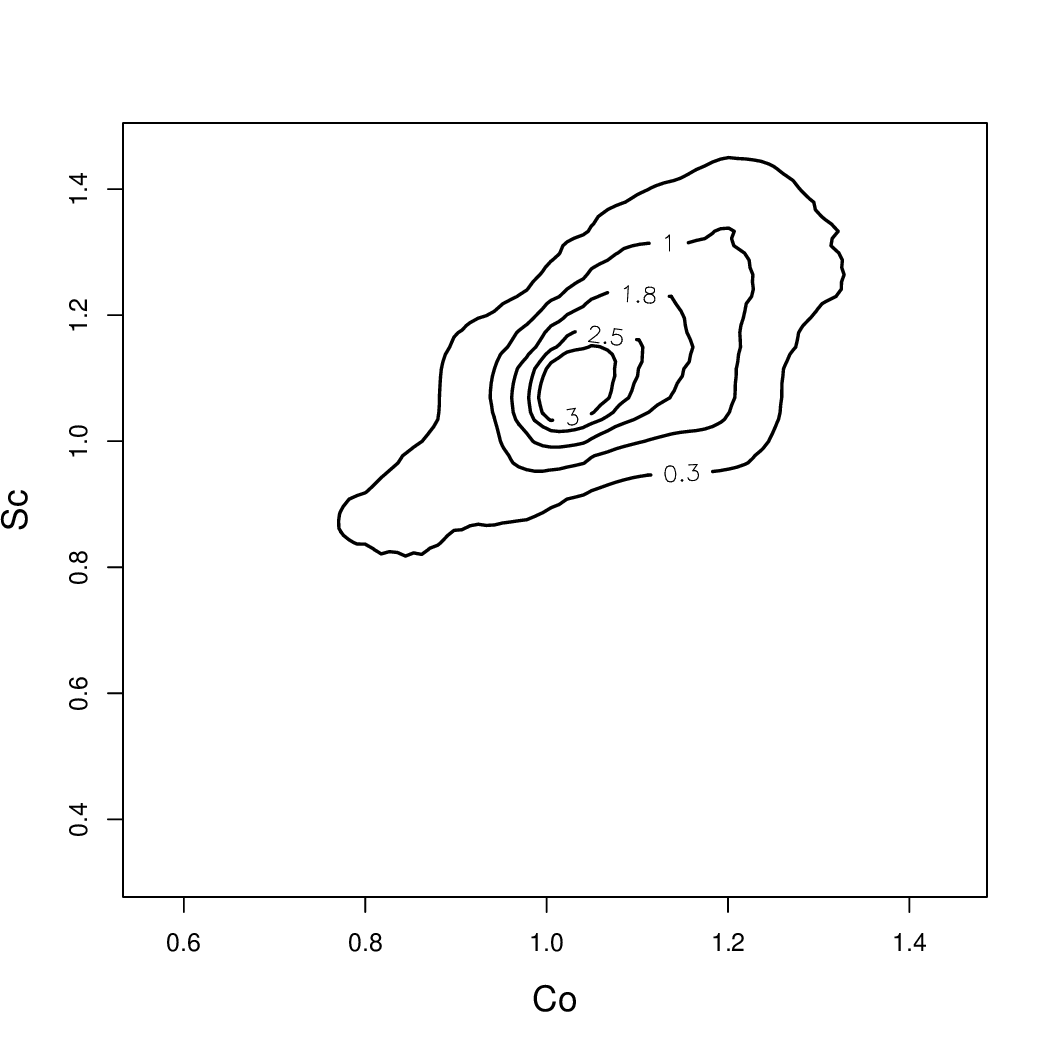} 
\end{tabular}
\end{center}
\vspace{-0.3cm}
\caption{
Scatter plots of (Co, Sc) for stratified values of Cs (first row), the estimated joint density contours with $\widehat{R}$ of sizes $5\times 5 \times 3$ (second row), $10\times 10 \times 5$ (third row), and $15\times 15 \times 3$ (fourth row). \newline
Left: Cs is small (Cs $\le 1.91$); Center: Cs is moderately-sized ($1.91 < $Cs $\le 2.13$); Right: Cs is large (Cs $> 2.13$).
}
\label{fig:uranium}
\end{figure}

We choose Cs as a stratifying variable and divide the data set into three parts
according to the level of Cs. 
The scatter plots and the joint density contours of Co and Sc with respect to different levels of Cs are shown in Figure \ref{fig:uranium}.
From the figure, we can see that the correlation structure of this data set is similar to
the structure of the simulation data in Section \ref{subsec:trivariate}
in that the correlation of the first two variables depends on the level of the third variable.
Specifically, the correlation between Co and Sc decreases when Cs increases.
Because the AIC tends to result in
small values of $(n_1,n_2,n_3) $ for $\widehat{R}$ and induces a poor fit to the observed data,
we investigate various choices of $(n_1,n_2,n_3) $ in estimating 
$\widehat{R}$ with the algorithm for 3-dimensional data.
Three examples of joint density contours with 
$(n_1,n_2,n_3) =(5,5,3)$, $(10,10,5)$ and $(15,15,3)$,
are shown in the last three rows of Figure \ref{fig:uranium}, respectively. 
Comparing the results, we can see that the estimated joint density contours 
with relatively larger sizes for $\widehat{R}$ fit the data more accurately.

\section{Discussion}
\label{sec:discussion}

We have developed EM algorithms to estimate the Bernstein copula 
for continuous or discrete data.
In this section, we provide some remarks for researchers who 
wish to use these algorithms
for practical data analysis, and we propose
further research topics.

In general, the convergence of the EM algorithm to the global maximizer
(i.e., the MLE) depends on the starting point of the algorithm.  In this 
paper, we propose to use the estimator given 
by \cite{Sancetta-Satchell04} and \cite{Janssen-etal12}, as 
stated in Remark \ref{rem:inconsistent}.
Numerical studies in Section \ref{subsec:consomic} provide empirical 
evidence that this starting point works well for calculating the MLE.
However, this result has not been established theoretically, and 
therefore it is essential to consider a variety of starting points 
for the algorithm.  

One of our primary objectives was to estimate the Bernstein copula.
Although the estimation of marginals was not considered in this paper,
it is worth noting that the estimation of marginals is of practical importance; 
for example, the ragged shape of the contour in Figure \ref{fig:Fmice}
disappears if we use a wider bandwidth in the kernel density estimation.

In Section \ref{subsec:consomic}, the AIC is used to choose
the size of the matrix $R=(r_{k,l})$. 
However, in our method, the marginal functions $F$ and $G$ are also unknown
and must be estimated, and for 
such semiparametric estimation procedures, the conventional AIC has no 
rationale.
In the construction of the conventional AIC, the asymptotic
distribution of the score function and its derivatives play crucial roles
\citep{Konishi-Kitagawa08}.  
\cite{Tsukahara05} developed a counterpart asymptotic theory
when the marginals are estimated
with empirical distribution functions; those results may be useful as building 
blocks for AIC methods for semiparametric copula estimation.
Also, we find in the example of Section \ref{subsec:uranium} that
the AIC has a tendency to choose smaller number of parameters
for this data set.
Theoretical and practical methods for selecting the model may be
a topic for further research.

An additional analysis of the data sets in Sections
\ref{subsec:consomic} and \ref{subsec:illinois}
shows that
the Bernstein copula and the Gaussian copula lead to similar results
for these data sets.
However, for data $(X,Y,Z)$ with 3-way interaction, such as
in Sections \ref{subsec:trivariate} and \ref{subsec:uranium},
we see that the Bernstein copula is more flexible for modeling correlation structures.
This is a remarkable feature not 
generally possessed by other copula families such as the Gaussian,
multivariate $t$-, and Archimedean copula families.
Recently, to describe such complicated data, the vine copula method
of creating a flexible copula by combining several copulas was developed
 \citep{Kurowicka-Joe11, Kim-etal13}.
The Bernstein copula is not only flexible in its own right,
but can also be incorporated into a vine copula 
to model data with more complicated correlation structures.
This is also a topic for further research.

\appendix
\section{Proof of Proposition \ref{prop:local_convergence}}

\newcommand{\m}{{\bm\mu}}
\renewcommand{\l}{{\bm\lambda}}
\newcommand{\f}{{\bm f}}
\newcommand{\g}{{\bm g}}
\renewcommand{\u}{{\bm u}}
\renewcommand{\v}{{\bm v}}
\newcommand{\0}{{\bf 0}}
\newcommand{\1}{\mbox{1}\hspace{-0.25em}\mbox{l}}

For vectors $\m=(\mu_k)_{1\le k\le m}$ and $\l=(\lambda_l)_{1\le l\le n}$,
define column vector valued functions by
\[
 \f(\m;\l) = (f_{k}(\m;\l))_{1\le k\le m}, \quad
 f_{k}(\m;\l) = \sum_{l=1}^n \frac{\bar\tau_{k,l}}{\mu_k + \lambda_l} - \frac{1}{m},
\]
and
\[
 \g(\l;\m) = (g_{l}(\l;\m))_{1\le l\le n}, \quad
 g_{l}(\l;\m) = \sum_{k=1}^m \frac{\bar\tau_{k,l}}{\mu_k + \lambda_l} - \frac{1}{n}.
\]
Let $\1_m=(\underbrace{1,\ldots,1}_m)'$.
Each step of Algorithm \ref{alg:m} can be rewritten as follows:

\smallskip

\begin{quote} 
Step M0.
Set the initial values for $\m^{(0)}=\m^{0}$.  Let $t=0$.

Step M1.
For fixed $\m^{(t)}$, let $\l^{(t)}$ be the solution of $\g(\l;\m^{(t)}) = \0$.

Step M2.
For fixed $\l^{(t)}$, let $\widetilde\m^{(t+1)}$ be the solution of
$\f(\m;\l^{(t)}) = \0 $.

Step M3.
Let
\[
 \m^{(t+1)} = \widetilde\m^{(t+1)}
 - \frac{1}{m}\1_m\1_m'(\widetilde\m^{(t+1)}-\m^{(0)})
\]
so that $\sum_k (\m^{(t+1)})_k = \sum_k (\m^{(0)})_k$.

Let $t:=t+1$ and go to Step M1, until $\m^{(t)}$ and $\l^{(t)}$ converge.
\end{quote}

Let $(\m^{*},\l^{*})$ be a solution of
$\g(\l^{*};\m^{*}) = \0$, $\f(\m^{*};\l^{*}) = \0$.
Since $(\m^{*}+r\1_m,\l^{*}-r\1_n)$ is also a solution
for arbitrary $r\in\mathbb{R}$,
we assume without loss of generality that
$\sum_k (\m^{0})_k = \sum_k (\m^{*})_k$.
From Step M1, we obtain 
\begin{align*}
\0 \
& = \g(\l^{(t)};\m^{(t)}) \\
& \doteq \g(\l^{*};\m^{*})
 + \nabla_\lambda \g(\l^{*};\m^{*})(\l^{(t)}-\l^{*})
 + \nabla_\mu \g(\l^{*};\m^{*}) (\m^{(t)}-\m^{*}), 
\end{align*}
and hence
\begin{equation} 
\l^{(t)}-\l^{*} \doteq -(\nabla_\lambda \g(\l^{*};\m^{*}))^{-1} \nabla_\mu \g(\l^{*};\m^{*}) (\m^{(t)}-\m^{*}).
\label{eqg}
\end{equation}
Here `$\doteq$' means that the difference of left-hand side and right-hand side
is of the order
$o\bigl(\max(\Vert\m^{(t)}-\m^{*}\Vert,\Vert\l^{(t)}-\l^{*}\Vert)\bigr)$.
Similarly, from Step M2,
\begin{align*}
\0 \
& = \f(\widetilde\m^{(t+1)};\l^{(t)}) \nonumber \\
& \doteq \f(\m^{*}; \l^{*})
 + \nabla_\mu \f(\m^{*};\l^{*})(\widetilde\m^{(t+1)}-\m^{*})
 + \nabla_\lambda \f(\m^{*};\l^{*}) (\l^{(t)}-\l^{*}),
\end{align*}
and hence
\begin{equation}
\widetilde\m^{(t+1)}-\m^{*}
\doteq -(\nabla_\mu \f(\m^{*};\l^{*}))^{-1}
 \nabla_\lambda \f(\m^{*};\l^{*}) (\l^{(t)}-\l^{*}).
\label{eqf}
\end{equation}
Because
$\1_m'\m^{(0)} = \sum_k (\m^{0})_k = \sum_k (\m^{*})_k = \1_m'\m^{*}$, 
we can rewrite Step M3 as
\begin{equation}
\m^{(t+1)} - \m^{*}
 = \widetilde\m^{(t+1)}
 - \frac{1}{m}\1_m\1_m'(\widetilde\m^{(t+1)}-\m^{*}) - \m^{*}
 = J (\widetilde\m^{(t+1)} - \m^{*}),
\label{eqJ}
\end{equation}
where
\[
 J = I_m - \frac{1}{m}\1_m\1_m'.
\]
Combining (\ref{eqf}) and (\ref{eqJ}),
we have
\begin{align}
\m^{(t+1)} - \m^{*} \
& = J (\widetilde\m^{(t+1)} - \m^{*}) \nonumber \\
& \doteq -J (\nabla_\mu \f(\m^{*};\l^{*}))^{-1}\nabla_\lambda \f(\m^{*};\l^{*}) (\l^{(t)}-\l^{*}).
\label{eqf'}
\end{align}

Let
$C=(c_{k,l})_{m\times n}$, 
$G=(\mathrm{diag}(c_{k+}))_{m\times m}$,
$H=(\mathrm{diag}(c_{+l}))_{n\times n}$, where
\[
 c_{k,l} = \frac{\bar\tau_{k,l}}{((\m^{*})_k+(\l^{*})_l)^2}, \quad
 c_{k+} = \sum_{l=1}^n c_{k,l}, \quad
 c_{+l} = \sum_{k=1}^m c_{k,l}.
\]
Simple calculations yield
\[
 -\nabla_\mu \f(\m^{*};\l^{*})
 = G, \qquad
 -\nabla_\lambda \g(\l^{*};\m^{*}) = H,
\]
\[
 -\nabla_\mu \g(\l^{*};\m^{*}) = C', \qquad
 -\nabla_\lambda \f(\m^{*};\l^{*}) 
 = C.
\]
Therefore, (\ref{eqg}) and (\ref{eqf'}) are rewritten as
\begin{align}
\l^{(t)}-\l^{*}   & \doteq -H^{-1} C' (\m^{(t)}-\m^{*}),  \label{update-y} \\
\m^{(t+1)}-\m^{*} & \doteq -J G^{-1} C (\l^{(t)}-\l^{*}). \label{update-x}
\end{align}
Combining (\ref{update-y}) and (\ref{update-x}), we have 
\begin{align}
\m^{(t+1)}-\m^{*}
&\doteq J G^{-1}CH^{-1}C' (\m^{(t)}-\m^{*}),
\label{mut-mu0}
\\
\l^{(t+1)}-\l^{*}
&\doteq H^{-1}C'J G^{-1}C (\l^{(t)}-\l^{*}).
\label{lambdat-lambda0}
\end{align}

To ascertain the asymptotic behavior
of $\m^{(t)}$ and $\l^{(t)}$ as $t\to\infty$,
we need to find the eigenvalues of the matrices
$J G^{-1}CH^{-1}C'$ and $H^{-1}C'J G^{-1}C$,
respectively \citep{Hageman-Young81}.
Let $D = G^{-1/2}CH^{-1/2}$; then we first show that the 
matrix $D$ has the largest singular value $\sigma_1(D)=1$.
This is because for column vectors $\u=(u_k)_{1\le k\le m}$ and
$\v=(v_l)_{1\le l\le n}$,
\[
 \u'C\v
 = \sum_{k,l} u_k v_l c_{k,l} 
 \le \sqrt{\sum_{k,l} u_k^2 c_{k,l} \sum_{k,l} v_l^2 c_{k,l}}
 = \sqrt{\sum_{k} u_k^2 c_{k+} \sum_{l} v_l^2 c_{+l}}
\]
and hence
\begin{align*}
\sigma_1(D)
&= \max_{\scriptsize\Vert\u\Vert=\Vert\v\Vert=1} \u'D\v
= \max_{\sum u_k^2=\sum v_l^2=1} \sum_{k,l} u_k v_l
 \frac{c_{k,l}}{\sqrt{c_{k+} c_{+l}}} \\
&= \max_{\sum u_k^2 c_{k+}=\sum v_l^2 c_{+l}=1} \sum_{k,l} u_k v_l c_{k,l} \\
&\le \max_{\sum u_k^2 c_{k+}=\sum v_l^2 c_{+l}=1}
 \sqrt{\sum_{k} u_k^2 c_{k+} \sum_{l} v_l^2 c_{+l}} = 1.
\end{align*}
This upper bound is attained when $\sigma_1(D)=\u_1'D\v_1$ with
\[
  \u_1 = \frac{1}{\sqrt{c_{++}}} G^{1/2}\1_m, \qquad
  \v_1 = \frac{1}{\sqrt{c_{++}}} H^{1/2}\1_n, \qquad
   c_{++} = \sum_{k,l}c_{k,l},
\]
because $\u_1'\u_1=\v_1'\v_1=1$ and
\[
 \u_1'D\v_1
 = \frac{1}{c_{++}}
(\1_m'G^{1/2}) G^{-1/2}CH^{-1/2} (H^{1/2}\1_n)
 = \frac{1}{c_{++}} \1_m'C\1_n =1.
\]
Therefore, $DD'$ has the largest eigenvalue 1,
and one of the corresponding eigenvectors is $\u_1$.

From the assumption that $\bar\tau_{k,l}>0$ for all $k,l$, we
find that $DD'$ is a positive matrix, i.e., all elements are positive.
By the Perron-Frobenius theorem \citep{Horn-Johnson90}, the 
multiplicity of the largest eigenvalue 1 of $DD'$ is 1, and so 
we obtain 
\[
 DD' = \frac{1}{c_{++}} G^{1/2}\1_m\1_m'G^{1/2}
 + \sum_{k=2}^m \nu_k \u_k \u_k',
\] 
where $1>\nu_2\ge\cdots\ge\nu_m\ge 0$ and
$0=\u_k'\u_1=\u_k'G^{1/2}\1_m/\sqrt{c_{++}}$.

Hence, the matrix $J G^{-1}CH^{-1}C'$ appearing in (\ref{mut-mu0})
is rewritten as
\begin{align}
J G^{-1}CH^{-1}C'
=&
J G^{-1/2} DD'G^{1/2} \nonumber \\
=& \frac{1}{c_{++}}
 J G^{-1/2} G^{1/2}\1_m\1_m'G^{1/2}G^{1/2}
 + J A = J A,
\label{JA}
\end{align}
where
\[
 A = G^{-1/2} \biggl( \sum_{k=2}^m \nu_k \u_k\u_k' \biggr) G^{1/2}.
\]
This matrix $JA$ has the same nonzero eigenvalues as those of
\[
AJ
 = G^{-1/2} \biggl( \sum_{k=2}^m \nu_k \u_k\u_k' \biggr) G^{1/2} \Bigl( I_m - \frac{1}{m}\1_m\1_m' \Bigr)
 = A,
\] 
which has the eigenvalues $\nu_2,\ldots,\nu_m$ and 0.
Here we used $0=\u_k'G^{1/2}\1_m$.
More precisely, it holds that
\[
 JA B = B N, \quad B=\bigl(\1_m,J G^{-1/2} (\u_2,\ldots,\u_m)\bigr),
\]
where $N=\mathrm{diag}(0,\nu_2,\ldots,\nu_m)$.
The matrix $B$ is nonsingular because
\[
 G^{1/2} B
  \begin{pmatrix}
   1  & \frac{1}{m}\1_m' G^{-1/2} (\u_2,\ldots,\u_m) \\
   \0 & I_{m-1}
  \end{pmatrix}
= \bigl(G^{1/2} \1_m,\u_2,\ldots,\u_m\bigr)
\]
is nonsingular.
Therefore,
\begin{equation}
 J A = B N B^{-1}.
\label{BNB1}
\end{equation}
Combining (\ref{BNB1}) with (\ref{mut-mu0}) and (\ref{JA}), we have
$B^{-1}(\m^{(t+1)}-\m^{*}) \doteq N B^{-1}(\m^{(t)}-\m^{*})$, and hence
for arbitrary $\varepsilon>0$
\begin{align}
 \Vert B^{-1}(\m^{(t+1)}-\m^{*})\Vert \le & \,(\nu_2+\varepsilon)\,
 \Vert B^{-1}(\m^{(t)}-\m^{*})\Vert
\label{nu2}
\end{align}
when $t$ is sufficiently large.

By proceeding in a similar manner, we find that the matrix $H^{-1}C'J G^{-1}C$
in (\ref{lambdat-lambda0}) can be diagonalized with the same eigenvalues
$1>\nu_2\ge\cdots\ge\nu_{\min(m,n)}\ge 0$ and 0 (if $m<n$).
Hence, an inequality of the same type as (\ref{nu2}) holds for the 
sequence $\l^{(t)}$.  
These inequalities imply the linear convergence of $\m^{(t)}$ and $\l^{(t)}$
with the rate $\nu_2+\varepsilon$.

\bigskip
\subsubsection*{Acknowledgments}

The authors thank the Illinois State Board of Education for permission to 
use the Illinois Standards Achievement Test scores data
analyzed in Section \ref{subsec:illinois}.
The authors are also grateful to Shingo Shirahata, Toshihiko Shiroishi and Toyoyuki Takada 
for their helpful comments.
This work was supported by the Systems Genetics Project
of the Transdisciplinary Research Integration Center,
Research Organization of Information and Systems.
The manuscript was partially presented
in the session on copulas of the 5th International Conference
of the ERCIM WG on Computing \& Statistics 
(ERCIM 2012, 1--3 December 2012, Oviedo, Spain), and
the initial manuscript was greatly improved by comments
provided by the participants.

\bigskip

\bibliographystyle{elsarticle-harv}

\end{document}